\documentclass[sigconf, nonacm, screen]{acmart}

\usepackage[utf8]{inputenc} 
\usepackage[T1]{fontenc}    
\usepackage{hyperref}       
\usepackage{url}            
\usepackage{booktabs}       
\usepackage{amsfonts}       
\usepackage{nicefrac}       
\usepackage{microtype}      
\usepackage{xcolor}         

\usepackage{algorithm}
\usepackage{algpseudocode}
\usepackage{hyperref}
\usepackage{epigraph}

\usepackage{soul}

\usepackage{graphicx, caption}
\usepackage{longtable}
\usepackage{geometry}
\usepackage{amsmath}
\usepackage[capitalise]{cleveref}
\usepackage{wrapfig}
\usepackage{multirow}
\usepackage{fancyvrb,fvextra}

\usepackage{lscape}
\usepackage{xspace}
\usepackage{booktabs}
\usepackage{tabularray}
\usepackage{codehigh}
\usepackage{dsfont}
\usepackage{array}
\usepackage{calc}
\usepackage{enumitem}
\usepackage{adjustbox}

\graphicspath{ {./images/} }

\usepackage{subcaption}
\usepackage{ifthen}

\newboolean{showComments}
\setboolean{showComments}{false}

\newcommand{\manish}[1]{
    \ifthenelse{\boolean{showComments}}
  {\textbf{\textcolor{blue}{[Manish: #1]}}}
  {}
}

\title{\textit{PhreshPhish}: A Real-World, High-Quality, Large-Scale Phishing Website Dataset and Benchmark}

\newcommand{\sharedaffil}{
  \affiliation{
    \institution{OpenText}
    \country{}
  }
}

\author{Thomas Dalton}
\authornote{Equal contribution.}
\sharedaffil
\email{tdalton@opentext.com}

\author{Hemanth Gowda}
\sharedaffil
\email{hgowda@opentext.com}

\author{Girish Rao}
\sharedaffil
\email{grao3@opentext.com}

\author{Sachin Pargi}
\sharedaffil
\email{spargi@opentext.com}

\author{Alireza Hadj Khodabakhshi}
\sharedaffil
\email{ahadjkhodaba@opentext.com}

\author{Joseph Rombs}
\sharedaffil
\email{jrombs@opentext.com}

\author{Stephan Jou}
\sharedaffil
\email{sjou@opentext.com}

\author{Manish Marwah}
\authornotemark[1]
\sharedaffil
\email{mmarwah@opentext.com}

\renewcommand{\shortauthors}{Dalton, Gowda, Rao, Pargi, Hadj Khodabakhshi, Rombs, Jou, and Marwah}

\usepackage{fancyhdr}

\fancypagestyle{firstpagepreprint}{
  \fancyhf{} 

  \fancyfoot[L]{\sffamily\small Preprint}
  
  \fancyfoot[C]{\thepage}

}

\begin{document}

\begin{abstract}

  Phishing remains a pervasive and growing threat, inflicting heavy
  economic and reputational damage. While machine learning has been
  effective in real-time detection of phishing attacks, progress is
  hindered by lack of large, high-quality datasets and benchmarks. In
  addition to poor-quality due to challenges in data collection,
  existing datasets suffer from leakage and unrealistic base rates,
  leading to overly optimistic performance results.
  In this paper, we introduce \textit{PhreshPhish}, a large-scale,
  high-quality dataset of phishing websites that addresses these
  limitations. Compared to existing public datasets,
  \textit{PhreshPhish} is substantially larger and provides
  significantly higher quality, as measured by the estimated rate of
  invalid or mislabeled data points.
  Additionally, we propose a comprehensive suite of
  benchmark datasets specifically designed for realistic model
  evaluation by minimizing leakage, increasing task difficulty,
  enhancing dataset diversity, and adjustment
  of base rates more likely to be seen in the real world.
  We train and evaluate multiple solution approaches to provide
  baseline performance on the benchmark sets. We believe the
  availability of this dataset and benchmarks will enable realistic,
  standardized model comparison and foster further advances in
  phishing detection. The datasets and benchmarks are available on
  Hugging Face (\url{https://huggingface.co/datasets/phreshphish/phreshphish}).

\end{abstract}

\maketitle
\thispagestyle{firstpagepreprint} 
\pagestyle{plain}

\section{Introduction}

Phishing attacks  are a pervasive and evolving threat that
uses social engineering techniques to deceive millions of
Internet users annually. According to the latest Phishing Activity
Trends Report \cite{apwg-report-2025q1} released by the Anti-Phishing Working
Group (APWG) \cite{apwg}, almost a million phishing attacks were observed
just in the first quarter of 2025. These attacks are a central component
of modern cyber kill chains \cite{cyberkillchain}, often involving
campaigns that lure users to spoofed websites designed to trick them into revealing
sensitive information, such as login credentials and financial data.
The data thus obtained often serves as an initial entry point to
fraud, identity theft and large-scale data breaches, resulting in substantial
financial and reputational damage. In 2024 alone, phishing contributed 
significantly to the estimated \$16.6 billion in cyberfraud losses
reported to the FBI \cite{fbi-ic3report-2024}.

Phishing remains a popular attack vector due to its effectiveness, low cost,
and ease of deployment. A major contributing factor is the widespread
availability of phishing kits \cite{cova2008there, bijmans2021catching}.
These kits contain complete phishing websites in a ready-to-deploy package, and
can often be downloaded for free or even purchased on the dark web markets \cite{dark-web-2018},
making such attacks accessible even to unskilled adversaries.

Despite considerable research efforts, detecting phishing sites
remains a challenging task \cite{das2019sok}. Various approaches have been proposed to
detect phishing, including use of blacklists, whitelists, heuristic-based
techniques and machine learning methods. However, its adversarial nature,
with attackers constantly adapting to avoid detection, and lack of good
quality data undermines these defense approaches. 
Phishing websites are typically short-lived, highly dynamic, and often
employ techniques such as obfuscation, cloaking \cite{9519414}, and
fast-flux hosting \cite{nagunwa2022machine}.

Deployed phishing detection systems can be integrated into various
applications, with browser plugins serving as a convenient option for
verifying websites, and blocking phishing sites before users access them.
To be effective in real-world settings, such systems must satisfy 
two key desiderata --- \textbf{(1) Low number of false postives:} legitimate
websites should only be rarely, if ever, blocked, implying a high precision,
$P(W|A)$\footnote{$W$ indicates a true phishing website; $A$ is a phishing website
identified by a model.},
where $1 - P(W|A)$ represents the proportion of incorrectly blocked sites;
and \textbf{(2) Low latency:} detection must operate in
near real time, ideally within a few hundred milliseconds, to minimize
user disruption.

\begin{figure*}
  \centering
  \includegraphics[width=\textwidth]{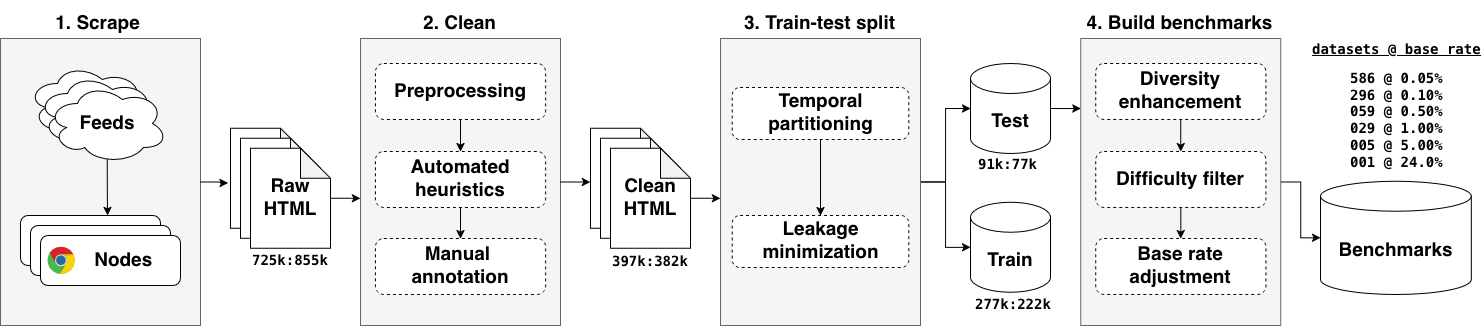}
  \caption{Our end-to-end pipeline consists of four distinct stages and we show the benign-to-phishing count at each step:
    (1) benign and phishing HTML is collected from the web using a real browser to ensure high fidelity;
    (2) the retrieved HTML is cleaned and assessed for quality using a combination of automated heuristics and human annotation;
    (3) train and test splits are created temporally and pruned to minimize leakage;
    (4) benchmark datasets are created by applying a set of diversity, difficulty and base rate filters.}
  \label{fig:pipeline}
\end{figure*}

State-of-the-art machine learning-based phishing detection systems
can be categorized into two broad approaches:
\textbf{(1) Direct phishing detection}, where models are trained to 
classify webpages directly as phishing or benign \cite{whittaker2010large,
  mohammad2015phishing, shahrivari2020phishing, aljofey2022effective}.
These models rely on feature extraction or representation learning from
a (partially) labeled dataset of webpages.
\textbf{(2) Target brand-based phishing detection}, where models are
trained to identify the brand a webpage attempts to mimic, and
phishing is inferred from inconsistencies between the webpage's URL and
the legitimate URLs associated with the identified brand \cite{afroz2011phishzoo, lin2021phishpedia, li2024phishintel}.
While the latter approach does not require a dataset of phishing 
webpages, it relies on a curated set of target brands, and
typically incurs much higher detection latency.

While the collection of large-scale web datasets is itself nontrivial, 
successfully scraping phishing pages, which behave adversarially to avoid
detection, presents a unique set of challenges beyond those typically encountered in
general web scraping. In addition to the fact that the modern web is increasingly 
\textbf{dynamic}, issues such as \textbf{ephemerality}, \textbf{cloaking}, and 
\textbf{staleness} significantly complicate the collection of high-quality datasets which
may be used to train and evaluate phishing detection models.
These challenges necessitate the development of robust methodologies for collecting, cleaning,
and validating phishing datasets to ensure their quality and utility for research purposes.
We believe that these obstacles, discussed in greater detail in Section~\ref{sec:challenges},
contribute to the overall lack of high-quality datasets in this area.
Although a search for publicly available phishing datasets would appear to yield promising
results, we observed substantial quality issues with many datasets (see Section~\ref{sec:cleaning-process}).
We also could not find any suitable benchmark datasets that facilitated a robust 
comparative evaluation of detection techniques despite the significant role that
phishing plays in cybercrime.

In this work, we present {\em PhreshPhish}, a large-scale, high-quality
phishing website dataset collected over more than one year.
To the best of our knowledge, this is the largest publicly available
phishing website dataset and offers significantly
higher quality compared to other such public datasets.
Further, we create realistic benchmark datasets for comparison between detection
methods. Figure~\ref{fig:pipeline} provides a schematic of the entire data pipeline from
obtaining the phishing and benign URLs to finally producing the training,
test and benchmark datasets. To summarize, we  make the following key contributions:

\begin{itemize}[topsep=4pt]
\item Publicly release {\em PhreshPhish}, a large-scale, high-quality phishing
  website dataset, that addresses drawbacks of existing such datasets. 
  Unlike prior datasets, our dataset is (1) larger-scale; (2) higher quality; (3) more recent (with
  planned periodic versioned updates).

\item We also release a suite of benchmark datasets to standardize the evaluation of detection methods.
  These benchmarks span five different base rates, ranging from 0.05\% to 5\%, making them
  more representative of real-world settings; further, we employ filters to increase dataset
  difficulty and diversity, and minimize data leakage.

\item We evaluate and compare a number of baseline methods on our benchmark dataset 
  to demonstrate the utility of our dataset and benchmarks.
  We also provide a detailed analysis of the performance of these methods.
  
\end{itemize}

\section{Related Work}

\subsection{Phishing Detection}



%

Phishing attacks span multiple channels, including email, SMS text messages (smishing), and social media \cite{aleroud2017phishing}.
Detecting phishing at its point of origin (e.g., in email or SMS) can reduce exposure early in the attack lifecycle, but it typically requires channel-specific solutions and new channels such as voice (vishing) continue to emerge \cite{enisa2025threatLandscape}.
A common downstream goal across channels is the harvesting of user credentials, which is often accomplished via a webpage.
Accordingly, we focus our discussion on \textit{webpage}-based phishing detection systems, which are commonly deployed either (1) in real-time to prevent users from accessing harmful sites, or (2) offline to support blacklist/whitelist generation.
The former imposes strict latency constraints, making computationally intensive methods or those requiring features that are slow to retrieve unsuitable for use.

Webpage phishing detection approaches have evolved over time
\cite{tang2021survey, zieni2023phishing, li2024state}, with early ones based on blacklists and whitelists
of URLs. Although fast and effective for known sites, this approach
suffers from complexity of maintaining these dynamic lists and its
inability to detect zero-day phishing attacks. Subsequent methods
used rules and heuristics based on features, which were later
leveraged for machine learning models. The features can be broadly
categorized into three types: (1) URL-based, (2) HTML-based, and (3)
host-based.

URL-based features refer to lexical features derived from the entire
URL or its components \cite{ma2009beyond, ma2011learning}. Since URLs
contain strong discriminative signals and are relatively short, a
large number of datasets and methods focus solely on URLs-based
features for detection \cite{el2020depth}.
HTML-based features are extracted from the HTML of a page, including
its content, structure (DOM tree), tags, embedded links, and
its visual rendering \cite{mohammad2015phishing}.
Host-based features relate to \textit{where} a website is hosted and
include its IP address, domain name properties, geographic information,
and similar attributes\cite{ma2009beyond}. Due to their retrieval
latency, these features are less suitable for real-time detection
scenarios, which typically require response times under a few hundred
milliseconds. Since our dataset is focused more towards real-time
detection, it excludes host-based features.

A wide range of machine learning approaches have been applied to
phishing detection, spanning from hand-crafted features to learning
representations of raw inputs. While many methods aim to directly
classify phishing pages \cite{whittaker2010large}, others have focused on identifying the impersonated
brand first, usually using webpage snapshots and a reference dataset of
known brands, followed by checking URL consistency \cite{afroz2011phishzoo, abdelnabi2020visualphishnet, lin2021phishpedia, li2024phishintel}.
Most recently, LLMs are being used for detecting phishing pages \cite{li2024knowphish,
  koide2024chatphishdetector, cao2025phishagent, zhang2025benchmarking}.
LLMs now have large enough context sizes that an entire HTML page
can fit. Although they are computationally expensive, large datasets
or model training is not required. 

\subsection{Existing Datasets}

Because many phishing detection systems are often focused on delivery mechanism (e.g., email, SMS, etc.), an internet search for phishing datasets appears to yield many results.
However, when filtering for datasets suitable for training and evaluating \textit{webpage}-based phishing detection systems, the results are much more limited. 
We searched for the terms \textit{"phishing"} and \textit{"phish"} on several common dataset hosting sites, including Hugging Face and Kaggle.
We excluded smaller datasets -- e.g., less than 10K rows on HuggingFace.
Almost all datasets we found were focused on email-oriented phishing detection, with only a few datasets containing URL and/or HTML data.

\begin{table}[ht]
  \centering
  \small
  \caption{Phishing datasets on common hosting sites.}
  \label{tbl:phishing-datasets}
  \begin{tabular}{@{}l c *{3}{c}@{}}
    \toprule
    Repository      & Total & \multicolumn{3}{c}{Feature type}      \\
    \cmidrule(lr){3-5}
                    &       & URL only & Email/SMS      & \textbf{URL \& HTML} \\
    \midrule
    Hugging Face    & 35    & 15       & 14         & 2                     \\
    Kaggle          & 19    & 6        & 4          & 4                     \\
    Harvard Dataverse & 2   & 0        & 2          & 0                     \\
    OpenML          & 4     & 2        & 1          & 0                     \\
    UCI ML Repo     & 3     & 3        & 0          & 0                     \\
    \bottomrule
  \end{tabular}
\end{table}

Four datasets on Kaggle \cite{kaggle-huntingdata11, kaggle-haozhang1579,
  kaggle-jackcavar, kaggle-aljofey}
and two on Hugging Face \cite{huggingface-ealvaradob, huggingface-huynq3Cyradar}
include raw HTML, however, these datasets are either small, outdated, or exhibit
substantial quality issues (see Section~\ref{sec:cleaning-process} for quality analysis).
Additionally, we found most of the other datasets lacked standardized test or benchmark splits,
and where splits were provided they were split randomly, which can lead to
data leakage and over-optimistic performance estimates.







\subsection{Minimizing Data Leakage}


Most approaches to splitting a dataset into a training and a test set
assume the data to be i.i.d. and perform a random split. A common
exception is time-series data, where ``future'' data points are usually used
for testing. However, data leakage \cite{kaufman2012leakage} can still
occur due to other factors. Kapoor et al. \cite{kapoor2022leakage}
provide a taxonomy of data leakage in machine learning, including
cases where training and test data are not independent. A typical
example is when multiple data points are collected from the same
entity, such as a patient \cite{oner2020training}, where random
splitting can place data from the same entity in both sets, causing
leakage. When entities are easily identifiable, data can be split to
avoid overlap. In phishing detection, one potential source of leakage
is sites built using the same kit.
However, since such information is usually not
available, we adopt a simple approach: we directly compare test and
training points for similarity, dropping test points that are too
similar to any training point. To make this search scalable to a large
dataset, we use locality sensitive hashing \cite{andoni2008near}.

\section{PhreshPhish Dataset}
\label{sec:phreshphish}




We collected our dataset over 17 months (July 2024--December 2025), encompassing a broad range of phishing and benign URLs.
Phishing URLs were primarily sourced from PhishTank \cite{phishtank}, the AntiPhishing Working Group (APWG) eCrime eXchange \cite{apwgecrimex}, and NetCraft \cite{netcraft}.
Benign URLs were drawn from anonymized browsing telemetry from over six million global Webroot users and Google search results for heavily targeted brands.
These user-sourced URLs provide a more realistic and representative sample of benign pages that users are likely to encounter in practice, relative to benign URLs drawn from web-crawl datasets such as Common Crawl \cite{commoncrawl}.
The dataset was cleaned and curated to reduce common sources of noise and labeling artifacts observed in prior datasets.

\subsection{Scraping Challenges}
\label{sec:challenges}

Web scraping is a standard component of many data collection pipelines designed to collect large training corpora for domains ranging from natural language processing to computer vision. 
However, data collection for applications in computer security remains difficult, particularly in the phishing domain, due to the adversarial nature of the setting.
In this section we identify common challenges encountered when collecting phishing datasets and describe the mitigation strategies incorporated into our pipeline.
We posit that these factors contribute to the scarcity of large, high-quality datasets in the phishing literature.

\begin{table}[h]
    \centering
    \small
    \caption{Common failure modes encountered when scraping phishing webpages.}
    \label{tab:failure-modes}
    \setlength{\tabcolsep}{4pt}
    \begin{tabular}{p{0.33\columnwidth} p{0.61\columnwidth}}  
        \toprule
        \textbf{Failure Mode} & \textbf{Description} \\
        \midrule
        Takedown Notice        & Host (e.g., Cloudflare) has removed the page and displays a standard "page removed" or takedown banner. \\
        Cloaking               & Malicious domain now redirects to a benign or unrelated page, sometimes to avoid detection. \\
        CAPTCHA                & Page requires solving a CAPTCHA, preventing automated scraping.
                                 This primarily applies to legitimate pages but is increasingly used by phishing pages as a cloaking strategy. \\
        JavaScript-Heavy Page  & Critical content is loaded dynamically via JavaScript and is not captured by traditional crawlers. \\
        HTTP Errors (4xx/5xx)  & Page returns error codes like 404 (Not Found), 403 (Forbidden), or 500 (Internal Server Error). \\
        Geofencing or IP Block & Page returns alternate content or blocks access based on the IP region or blacklists scrapers. \\
        \bottomrule
    \end{tabular}
\end{table}

\begin{figure*}[ht]
  \centering
  \begin{subfigure}[c]{0.50\textwidth}
    \centering
    \includegraphics[width=\textwidth]{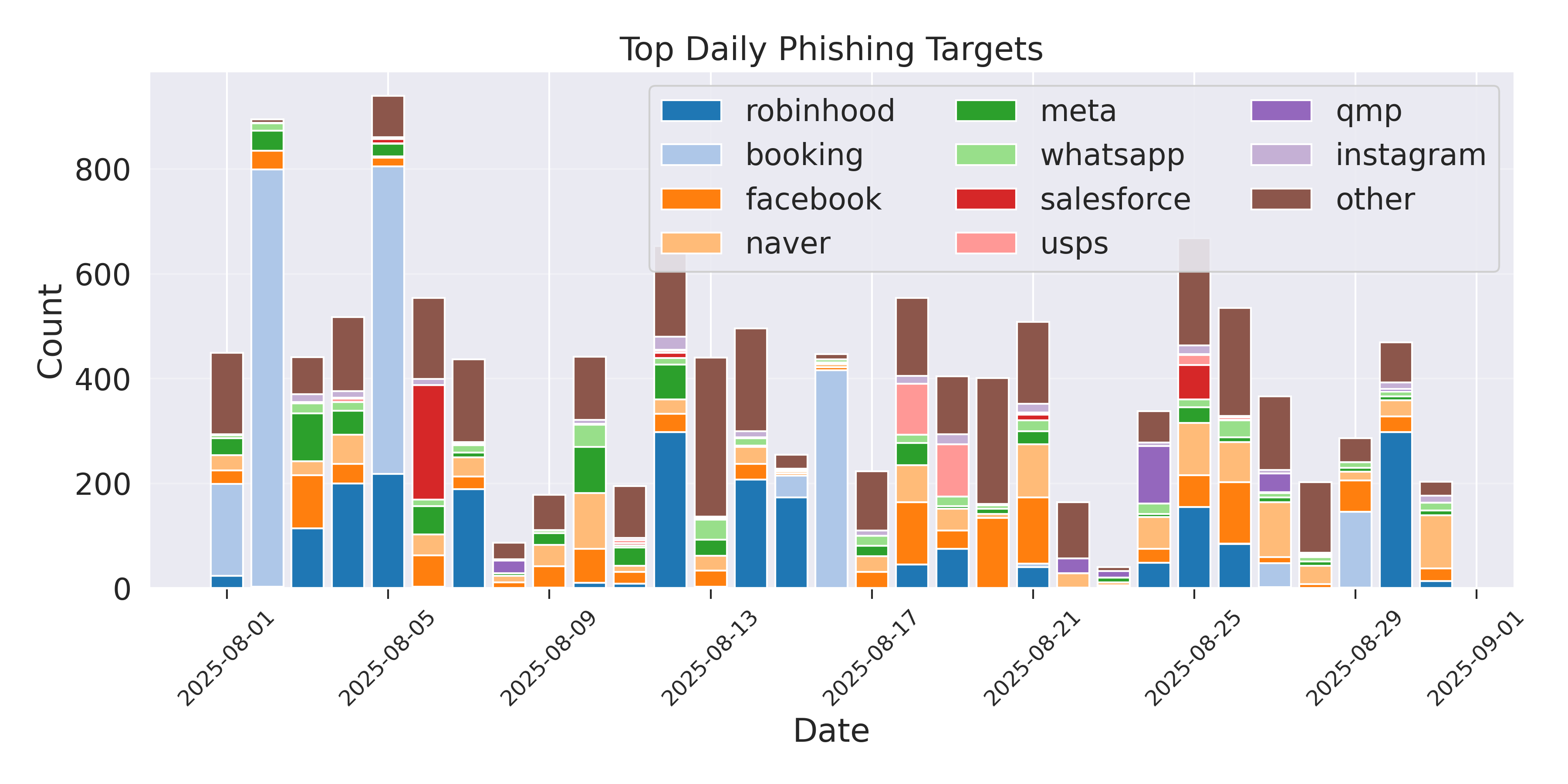}
    \caption{\label{fig:ourdataset:top_targets_daily}}
  \end{subfigure}
  \hfill
  \begin{subfigure}[c]{0.43\textwidth}
    \centering
      \includegraphics[width=\textwidth]{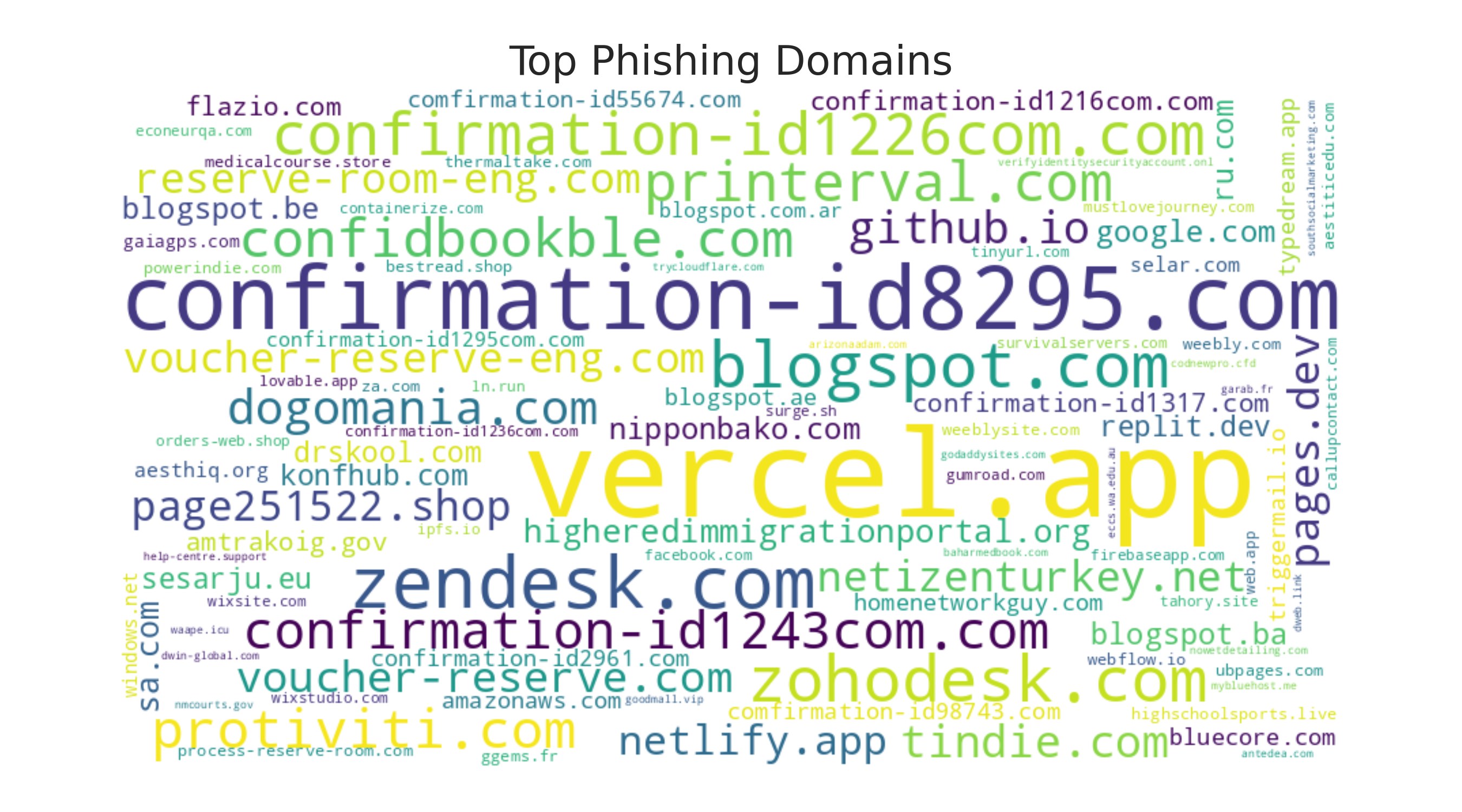}
    \caption{\label{fig:ourdataset:word_cloud}}
  \end{subfigure}

  \caption{A one-month snapshot (August 2025) of collected phishing pages. (\subref{fig:ourdataset:top_targets_daily}) Most frequently targeted brands over time.
  Targeted brands follow a power law-like distribution, with a few brands being targeted much more frequently than others.
  (\subref{fig:ourdataset:word_cloud}) The most common domains in phishing URLs are often legitimate domains that allow users to upload and host content such as Vercel and Blogspot.}
  \label{fig:ourdataset:combined}
\end{figure*}

\textbf{Dynamic pages}
Modern websites rely on JavaScript and frameworks like React \cite{react} and Angular \cite{angular} to load content dynamically.
This poses a challenge for traditional crawlers, which only fetch static HTML and cannot execute client-side scripts.
Consequently, dynamic content is often missed, limiting the usefulness of datasets such as Common Crawl and datasets derived from it \cite{raffel2020exploring,penedo2024fineweb,datacommons}, which rely on traditional scraping tools like Apache Nutch \cite{nutch}.

\textbf{Ephemerality}
Phishing pages are often short-lived and may be taken down before they can be scraped.
This is particularly true for pages that are hosted on compromised benign domains, which are often taken down by the domain owner or hosting provider.
This can lead to situations where the scraper is unable to access the original phishing page and instead is presented with a benign page or a takedown notice.
Datasets that do not address this challenge risk containing significant amounts of mislabeled data.

\textbf{Cloaking}
Cloaking \cite{9519414} is a technique used by attackers to present different content conditioned on specific criteria, such as a client's IP address or user agent.
This can lead to situations where the scraper is presented with a benign page instead of the phishing page, resulting in mislabeled data.
This adversarial technique can lead to model poisoning \cite{tian2022comprehensive} if the mislabeled data is used for training, and can lead to over-optimistic performance estimates if the mislabeled data is used for evaluation.

\textbf{Staleness}
Phishing techniques and tactics constantly evolve, with new pages and campaigns emerging regularly.
In addition, underlying web technologies change over time, with new standards like HTML5 and frameworks like GraphQL \cite{graphql} being adopted.
Although concept drift is a challenge for all machine learning applications, it is particularly acute in phishing detection where evasion tactics evolve rapidly.
Many existing phishing datasets are outdated and no longer relevant, as they were collected years ago and do not reflect the current state of the web.
To combat the issue of staleness, we plan to maintain \textit{PhreshPhish} with regular updates to ensure its relevance and utility.

To address the challenges outlined above, we developed a scraping pipeline that is designed to maximize the probability of capturing high-quality phishing data.
Our scraping pipeline leverages Selenium \cite{selenium} to render pages in full browser instances, capturing dynamic content missed by traditional tools.
Pages are scraped using a distributed cluster running Windows and Chrome, allowing us to address the challenges of dynamic pages and cloaking.
To further defend against cloaking, nodes in the cluster are configured to use a variety of user agents and IP addresses.
To mitigate the ephemerality challenge, URLs are scraped within minutes of being reported to our feeds, maximizing the likelihood of capturing the original phishing page before it is taken down.
Finally, we continuously monitor the performance of our scraping system and make adjustments as needed to ensure that it remains effective.

\subsection{Cleaning Process}
\label{sec:cleaning-process}
Although our scraping system is designed to maximize the probability of capturing live phishing pages, scrape failures corresponding to the failure modes outlined in Section~\ref{sec:challenges} can still occur.
Therefore, we developed a cleaning process designed to remove such failures from the dataset.
This cleaning process consists of two stages: a heuristics-based automated filtering stage and a human-in-the-loop manual inspection stage.

\textbf{Automated filtering}
Because URLs are often reported to multiple feeds, we normalize the URL to a canonical form and deduplicate the data at the URL level.
The canonicalization process includes removing trailing slashes and converting the URL to lowercase
\footnote{
    Although a URL can contain uppercase letters, we convert it to lowercase only for deduplication purposes since it is often the case that the same URL will be reported multiple times, occasionally with different casing.
    After deduplication, the original case-sensitive URL is retained.
}.
Next, the HTML is parsed and the title of the page is extracted.
Pages with titles containing certain keywords such as \textit{"forbidden"} or \textit{"not found"} are removed as they are likely to be scrape failures.

\textbf{Manual inspection}
After automated filtering, pages are then grouped together and a representative page, known as a prototype, is selected from each group for manual inspection.
A human annotator inspects the prototypical page and decides whether to keep or remove the page from the dataset.
The decision is then applied to all nearest neighbors of the prototype with respect to cosine similarity of the TF-IDF features of the page HTML.
This process continues until a specified budget (i.e. the maximum number of prototypes to inspect) is exhausted.
To maximize the efficiency of the cleaning process, we use two distinct grouping schemes.
The first grouping is done by extracting TF-IDF features from the page HTML and performing locality-sensitive hashing (LSH) via random projection to group similar pages together.
The second grouping is done with respect to page title.
For each scheme, the groupings are then ordered from largest to smallest with larger groups being inspected first.
Although the cleaning stage requires manual human inspection and annotation, it is optimized to an extent due to the fact that the human annotator only needs to inspect a representative page from each group and any decision made about the prototypical page is applied to all nearest neighbors in bulk.
This significantly reduces the amount of manual work required to clean the dataset.
Furthermore, a budget is imposed on the number of prototypes to be inspected to ensure that the cleaning process is efficient and does not take an excessive amount of time.
In practice, this manual stage is lightweight; a typical annotation budget is 1--3 hours.

Finally, pages likely to contain personally identifiable information (PII) are also removed from the dataset.
Full details including pseudocode for cleaning and our PII removal process are provided in Appendix~\ref{sec:appendix-dataset}.

\subsection{Quality Assessment}
We next assess dataset quality by comparing \textit{PhreshPhish} to two prominent datasets in the literature: Aljofey, et al. \cite{aljofey2022effective} and Crawling2024 \cite{xie2025scalable}.
We apply the same cleaning methodology used to construct our dataset (Section~\ref{sec:cleaning-process}) and use the observed rejection rates as a proxy for dataset quality.
Because complete manual inspection is not feasible, we allocate a manual cleaning budget of approximately one hour of annotator time per dataset (60 LSH bins and 50 title groups).
Then we estimate a lower and upper bound for the amount of data remaining if cleaning were to continue to completion.
The results of our quality assessment are summarized in Table~\ref{tab:quality-assessment}. 

\begin{table*}[t]
  \centering
  \small
  \caption{Our two-stage cleaning pipeline applied to three phishing datasets.}
  \label{tab:quality-assessment}
  \renewcommand{\arraystretch}{0.82}
  \setlength{\tabcolsep}{4pt}
  \begin{tabular*}{\textwidth}{@{\extracolsep{\fill}}l rrr rrr rrr@{}}
    \toprule
    & 
    \multicolumn{3}{c}{\textbf{Aljofey, et al.}} & 
    \multicolumn{3}{c}{\textbf{Crawling2024}} & 
    \multicolumn{3}{c}{\textbf{PhreshPhish (ours)}} \\
    & Phish & Benign & Total & Phish & Benign & Total & Phish & Benign & Total \\
    \midrule
    \multicolumn{10}{l}{\textit{Stage 0: Initial Size}} \\
    Initial Size ($N$)          & 27,280 & 32,972 & 60,252 & 26,687 & 29,428 & 56,115 & 250,002 & 366,991 & 616,993 \\
    \midrule
    \multicolumn{10}{l}{\textit{Stage 1: Automated Filtering}} \\
    URL Duplicates            & 2,936  & 129    & 3,065  & 1,069  & 166    & 1,235  & 0       & 0       & 0       \\
    HTML Missing              & 1,843  & 1,103  & 2,946  & 1,427  & 103    & 1,530  & 0       & 0       & 0       \\
    Bad Title                 & 12,715 & 5,018  & 17,733 & 6,322  & 665    & 6,987  & 0       & 0       & 0       \\
    \textbf{Remaining after Stage 1}    & 12,722 & 26,851 & 39,573 & 18,938 & 28,660 & 47,598 & 250,002 & 366,991 & 616,993 \\
    \midrule
    \multicolumn{10}{l}{\textit{Stage 2: Manual Inspection}} \\
    LSH Rejections            & 2,530  & 5,422  & 7,952  & 3,390  & 3,470  & 6,860  & 1,927   & 790     & 2,717   \\
    Title Rejections          & 794    & 183    & 977    & 555    & 47     & 602    & 491     & 0      & 491      \\
    \textbf{Remaining after Stage 2}    & 9,398  & 21,246 & 30,644 & 14,993 & 25,143 & 40,136 & 247,584 & 366,201 & 613,785 \\
    \midrule
    \textbf{Total Removed}    & 17,882 & 11,726 & 29,608 & 11,694 & 4,285  & 15,979 & 2,418   & 790     & 3,208   \\
    \textbf{Rejection Rate} ($r$)   & 65.5\% & 35.6\% & 49.1\% & 43.8\% & 14.6\% & 28.5\% & 0.97\%  & 0.02\%  & 0.52\% \\
    \bottomrule
  \end{tabular*}
\end{table*}

\textbf{Aljofey, et al. \cite{aljofey2022effective}}
This dataset contains 60k samples, with 27k labeled as phishing and 33k labeled benign.
We found this dataset contained a large number of data points that could immediately be removed under our automated filtering process (the majority of which were due to having bad titles such as "404 Not Found").
After running our cleaning process, over 65\% of the phishing samples in this dataset were removed, leaving only 9,398 phish samples remaining in the dataset.
The benign samples were also heavily affected, with nearly 36\% of the benign samples being removed.
Assuming the remaining un-inspected samples are perfectly clean, the dataset would contain 9,398 phishing samples and 21,246 benign samples.
Under more realistic assumptions and based on rejection rates observed, however, we estimate that the dataset would contain only around 3,242 phishing samples and 13,682 benign samples if cleaning were to continue to completion.

\textbf{Crawling2024 \cite{xie2025scalable}}
This dataset contains 56k samples, with 27k labeled as phishing and 29k labeled as benign.
Although this dataset performed better than Aljofey, et al., there were still significant quality issues with over 43\% of phishing samples being marked for removal during cleaning.
The dataset would contain 14,993 phishing samples and 25,143 benign samples for a total of 40,136 samples as a upper bound.
Under more realistic assumptions and based on rejection rates observed, however, we estimate that the dataset would contain only around 8,400 phishing samples and 20,000 benign samples if cleaning were to continue to completion.

\textbf{PhreshPhish (ours)}
Combining the train and full benchmark set, we have a total of 617k samples, with 250k labeled as phishing and 367k labeled as benign.
Because we already previously applied our cleaning process to the dataset, the automated filtering stage didn't result in any new samples being removed.
Upon manual inspection, fewer than 1\% of the phishing samples and a negligible number of benign samples were removed.
This is expected, and we do not claim that our dataset is free of all errors, as an unlimited cleaning budget is not feasible.
Nevertheless, the low rejection rate provides evidence of higher dataset quality relative to prior collections under the same auditing protocol.
Moreover, \textit{PhreshPhish} is substantially larger than existing datasets, with nearly 248k phishing samples and 366k benign samples remaining after cleaning (approximately an order of magnitude larger than the other datasets).

\section{Test and Benchmark Datasets}
\label{sec:test-and-benchmarks}

\begin{quote}
  \centerline{\em{For better or for worse, benchmarks shape a field.}}
   \hfill---David Patterson, 2017 Turing award winner \cite{patterson2012better}
\end{quote}

For model evaluation, we provide a test set and a suite of
benchmarks. The test dataset satisfies two key properties: (1)
it is temporally disjoint and lies in the future with respect to the
training data; and (2) it contains minimal leakage from the training
set. Despite these features, the test set is not fully representative of a
deployed real-world setting. In fact, as shown in Figure~\ref{fig:pr-bench}, 
the base rate has a significant impact on model performance. 
To support more realistic, standardized
evaluation and comparison of phishing models, we construct benchmark
datasets by applying three additional operations to the test dataset:
(1) difficulty filtering; (2) diversity enhancement; and (3) base rate
adjustment.

In contrast to our approach, we found most publicly available
phishing website datasets either lack a test or benchmark
dataset, or suffer from significant training-to-test set leakage,
leading to inflated performance evaluation. Further, the base rate
in most datasets is too high to serve as realistic benchmarks. 
For example,
Phishpedia \cite{lin2021phishpedia},
ChatPhishDetector \cite{koide2024chatphishdetector},
and PhishAgent \cite{cao2025phishagent}
use a balanced test set (or 50\% base rate).
Others such as \cite{liu2023knowledge} and
\cite{liu2024less} also use similarly unrealistic base rates.

\subsection{Test Dataset}

Starting from the clean dataset, we first perform temporal
partitioning, followed by leakage filtering to remove test data points
that show high similarity to any point in the train set.
The pseudocode to generate the test set is shown in Algorithm~\ref{alg:test-set}
in Appendix~\ref{sec:appendix-bench}.

{\bf Temporal partitioning}
Although phishing data is not a time-series and no explicit temporal
dependence exists, phishing methods and campaigns at a particular time
show some degree of dependence and evolve over time.  To eliminate any
possibility of a model gaining useful information from future points
in the train set when evaluated on a test set, we partition the
dataset such that the test set lies strictly in the future relative to
the train set.

{\bf Leakage filter}
While temporal separation provides some protection against data leakage
from the train to the test set, 
very similar points could still exist between train and test datasets.
To mitigate this problem, we traverse the test set and discard points
whose TF-IDF cosine distance is less than a specified threshold from any train points.
LSH is used to make searching the train set more scalable. 
Figure~\ref{fig:dataset-temporal} in Appendix~\ref{sec:appendix-bench} shows the temporal
distribution of the train and test data sets based on date of
collection.

\subsection{PhreshPhish Benchmarks}

To encourage the development of more robust and generalizable phishing
detection models, we provide a suite of benchmark datasets. The benchmark
datasets are derived from the test dataset but are further processed to
make them more challenging and representative of real-world scenarios.
The steps to create the benchmarks are shown in Algorithm~\ref{alg:benchmark-set}
in Appendix~\ref{sec:appendix-bench}.

{\bf Diversity enhancement}
We prune as well as augment the test dataset to make it
more diverse and representative. 
We augment the benign samples with data points corresponding to the top
targeted brands to ensure representation of brands not captured
by our benign feed. 
The distribution of targeted brands is estimated across
the entire dataset, from which the $k$ most frequent brands are extracted.
We perform automated web searches to retreive URLs corresponding to these
brands, which are subsequently scraped and integrated into the test dataset.
For our dataset, we selected $k=173$, resulting in approximately 1,000 new 
data points after cleanup.

As part of diversity filter, we prune data points in the test set that
have similar URLs or HTMLs. This is applied to both phishing and benign points.
This is implemented similar to the leakage filter but points within the
test set are compared; any point with neighbors closer than a threshold is
dropped. 

{\bf Difficulty filter}
Applying the difficulty filter prunes out ``easy''
points, making the dataset more challenging to classify. There are
several approaches to discovering easy points including exploring various
data features, determining the most confounding feature values 
for the output classes, and selecting data points with those values.
We take a simpler approach and
assume the classification probability of a correctly
classified data point can be taken as a proxy of its 
difficulty. We run the test set through a classifier trained on the
training set. Correctly classified phishing points with score greater
than $1 - \delta$ are considered easy and candidates for pruning.
We randomly sample and drop $k\%$ of the entire dataset. Specifically,
we set $\delta = 0.15$, and $k = 10$.

{\bf Base rate adjustment}
The base rate of phishing points in our test set is 46\%. Other
publicly available test datasets have similar high base rates.
However, in practice the base rates a detector observes are much
lower, and vary depending on a multitude of factors including data
provenance and the model pipeline used in deployment.
To represent a wide range of scenarios, we
create a large number of benchmark datasets with varying base rates. In
particular, we create benchmark datasets with five different base rates varying
from 0.05\% to 5\%. In addition, we also provide the entire benchmark data set, which
has a base rate of 24\%.
To account for variance, multiple instances, $m$, are sampled for
each base rate, resulting in 975 total datasets.
For lower base rates, the number of phishing points in a benchmark set
are miniscule, and hence the variance is high.

\textit{Determining $m$:}
How many instances, $m$, are required, given a base rate, $b$, corresponding
to $k$ phishing points in a dataset of size $N$, for the model error estimate, $M_e$, to be 
accurate within a margin of error $e$ (e.g., $\pm 1\%$)? Since model error
is an average, we can assume
$M_e \sim \mathcal{N}(\mu_{M_e},\,\sigma_{M_e})$ (per central limit theorem).
$m$ is sufficiently large when
$m \geq \frac{z^2 \cdot \hat{\sigma_{M_e}}^2}{e^2}$
where $\hat{\sigma_{M_e}}^2$ is the empirical variance of the model
error and $z$ is the standard normal z-score corresponding to a desired
confidence level. 
For each $b$, we sample multiple small sets of instances to compute average
empirical variance, which is then used to estimate $m$. 
This analysis assumes sampling with replacement; sampling without replacement
generally requires fewer samples, however, the difference is negligible
when $k \ll N$, as is true in our case. 
Another advantage of sampling without replacement is that all available
points are covered.
In our experiments, we adopt sampling without replacement with $m$'s 
higher than the estimated lower bound with 1\% margin of
error and 95\% confidence level. The base rates and the corresponding
$m$'s in the benchmark datasets are: 
0.05\% (586), 0.1\% (296), 0.5\% (59), 1\% (29), 5\% (5).



\begin{figure*}[t]
  \centering

  \begin{subfigure}[t]{0.245\textwidth}
    \centering
    \includegraphics[width=\linewidth]{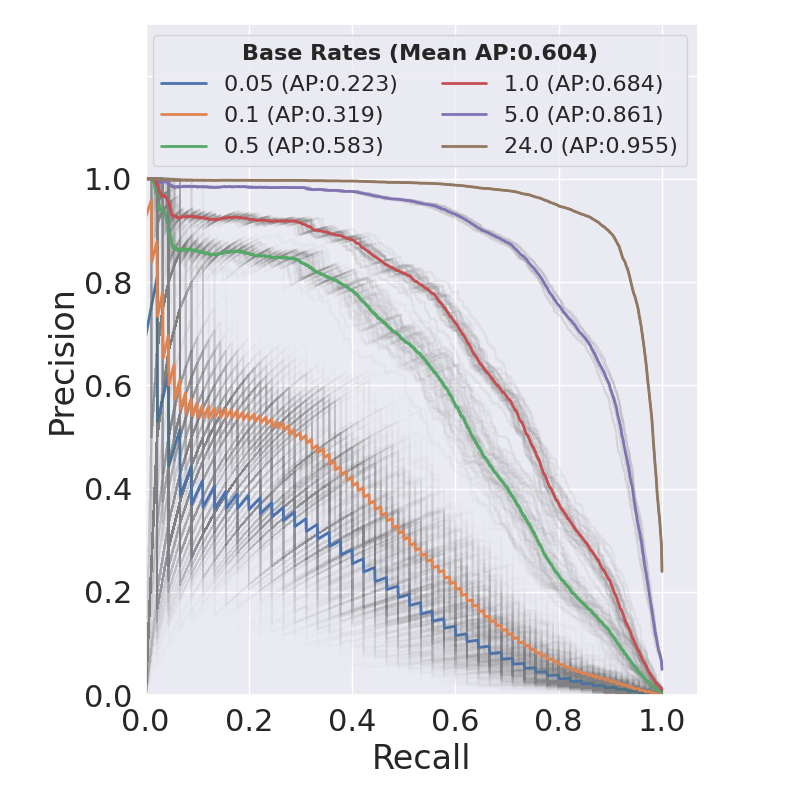}
    \caption{Linear model}
  \end{subfigure}\hfill
  \begin{subfigure}[t]{0.245\textwidth}
    \centering
    \includegraphics[width=\linewidth]{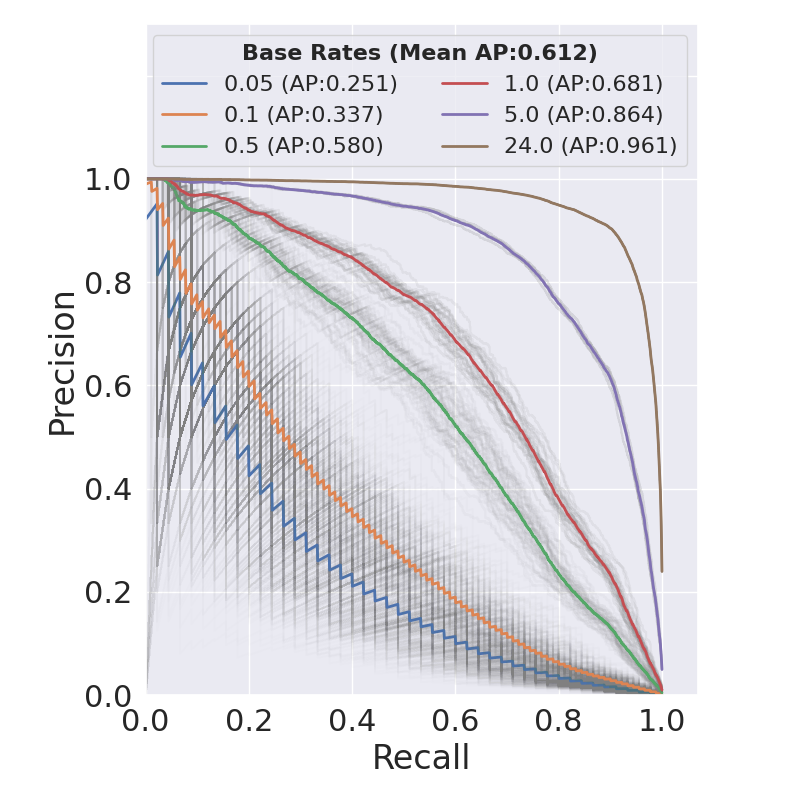}
    \caption{FFN model}
  \end{subfigure}\hfill
  \begin{subfigure}[t]{0.245\textwidth}
    \centering
    \includegraphics[width=\linewidth]{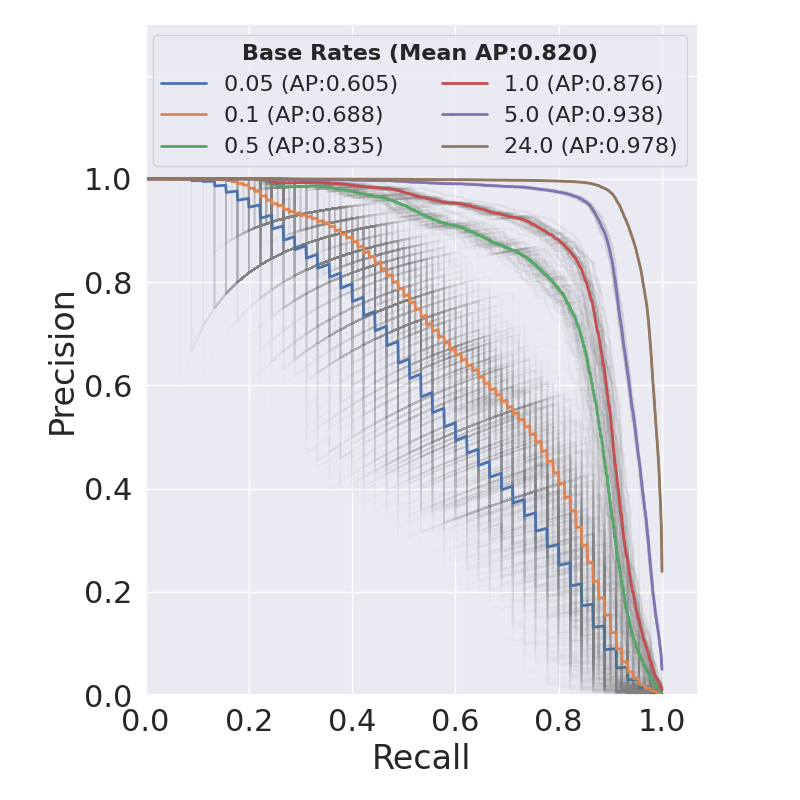}
    \caption{GTE model}
  \end{subfigure}\hfill
  \begin{subfigure}[t]{0.245\textwidth}
    \centering
    \includegraphics[width=\linewidth]{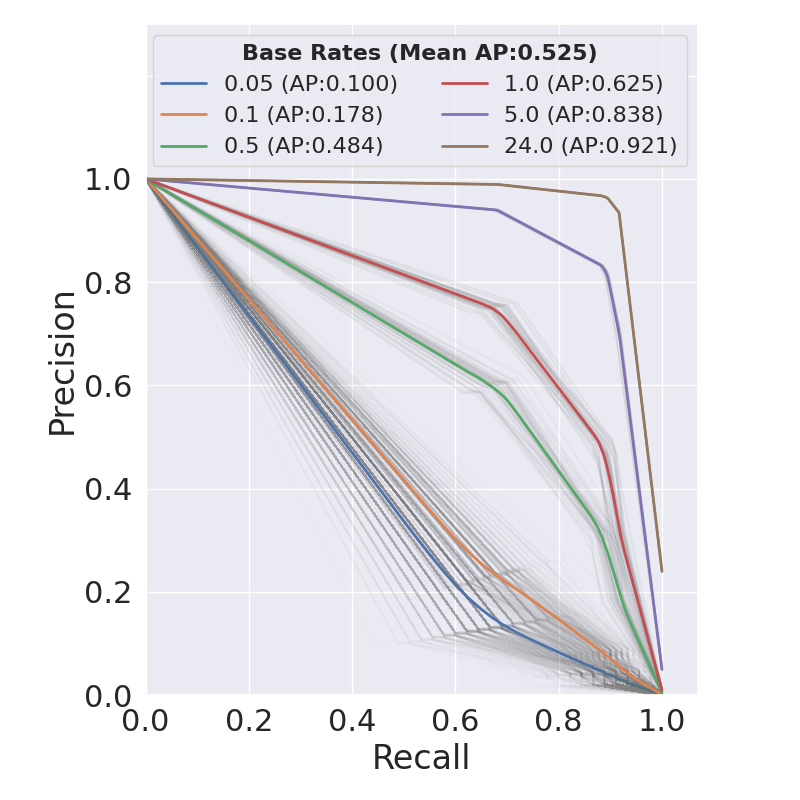}
    \caption{LLM}
  \end{subfigure}

  \caption{Precision-recall curves on the benchmark datasets}
  \label{fig:pr-bench}
\end{figure*}

As shown in Figure~\ref{fig:pr-bench}, model performance is
highly sensitive to the base rate, with performance improving as the
base rate increases. Importantly, most of the strong performance results
reported in the literature on this task can be largely attributed to
the use of unrealistically high base rates. While Figure~\ref{fig:pr-bench}
empirically shows the dependence of model precision on the base rate,
this relationship is supported by theory as well
(see Section~\ref{sec:baseRateDependence} in the Appendix).

\section{Baseline Models}
\label{sec:baselines}

Phishing detection is a binary classification problem where the input space $\mathcal{X}$ is 
the Cartesian product $\mathcal{X} = \mathcal{U} \times \mathcal{H}$ where $\mathcal{U}$ denotes the space of
URLs and $\mathcal{H}$ denotes the space of HTML content \footnote{
      Other input modalities such as host-based features (i.e. domain age, SSL certificate validity, etc.) could be used as well but we exclude them here since their inclusion tends to introduce latency that would be unacceptable for real-time detection.}.
The goal is to learn a function $f: \mathcal{U} \times \mathcal{H} \rightarrow [0,1]$ that estimates the conditional probability $P(Y=1|x),$ where $Y \in \{0,1\}$ is the binary class label (with $Y=1$ indicating a phishing page and $Y=0$ indicating a benign page) and $x=(u, h) \in \mathcal{X}$.
The function is learned from a labeled dataset $\mathcal{D} = \{(x_i,y_i)\}^N_{i=1}$.

\begin{table*}[ht]
\centering
\footnotesize
\resizebox{\textwidth}{!}{%
\begin{tabular}{l l *{6}{c}}
\toprule
\textbf{Model} &        & \textbf{0.05} & \textbf{0.10} & \textbf{0.50} & \textbf{1.0} & \textbf{5.0} & \textbf{24.0 (full)} \\
\midrule

\multirow{3}{*}{Linear}
& AP           & 0.2230 $\pm$ 0.0043 & 0.3191 $\pm$ 0.0047 & 0.5825 $\pm$ 0.0050 & 0.6841 $\pm$ 0.0045 & 0.8610 $\pm$ 0.0047 & 0.9551 $\pm$ 0.0028 \\
& P@R\,=\,0.90 & 0.0149 $\pm$ 0.0007 & 0.0298 $\pm$ 0.0014 & 0.1223 $\pm$ 0.0062 & 0.2191 $\pm$ 0.0102 & 0.5950 $\pm$ 0.0213 & 0.8957 $\pm$ 0.0075 \\
\midrule
\multirow{3}{*}{FFN}
& AP           & 0.2511 $\pm$ 0.0048 & 0.3368 $\pm$ 0.0050 & 0.5798 $\pm$ 0.0058 & 0.6813 $\pm$ 0.0047 & 0.8640 $\pm$ 0.0027 & 0.9607 $\pm$ 0.0018 \\
& P@R\,=\,0.90 & 0.0163 $\pm$ 0.0007 & 0.0316 $\pm$ 0.0012 & 0.1319 $\pm$ 0.0044 & 0.2339 $\pm$ 0.0068 & 0.6137 $\pm$ 0.0111 & 0.9045 $\pm$ 0.0038 \\
\midrule
\multirow{3}{*}{GTE}
& AP           & 0.6050 $\pm$ 0.0051 & 0.6879 $\pm$ 0.0044 & 0.8354 $\pm$ 0.0035 & 0.8762 $\pm$ 0.0032 & 0.9382 $\pm$ 0.0016 & 0.9779 $\pm$ 0.0020 \\
& P@R\,=\,0.90 & 0.0891 $\pm$ 0.0068 & 0.1563 $\pm$ 0.0123 & 0.3736 $\pm$ 0.0289 & 0.5426 $\pm$ 0.0305 & 0.8576 $\pm$ 0.0133 & 0.9746 $\pm$ 0.0037 \\
\midrule
\multirow{3}{*}{LLM}
& AP           & 0.1003 $\pm$ 0.0012 & 0.1780 $\pm$ 0.0020 & 0.4839 $\pm$ 0.0038 & 0.6252 $\pm$ 0.0044 & 0.8383 $\pm$ 0.0030 & 0.9216 $\pm$ 0.0034 \\
& P@R\,=\,0.90 & 0.0390 $\pm$ 0.0017 & 0.0722 $\pm$ 0.0035 & 0.2606 $\pm$ 0.0141 & 0.4147 $\pm$ 0.0214 & 0.7861 $\pm$ 0.0145 & 0.9352 $\pm$ 0.0235 \\

\bottomrule
\end{tabular}%
}
\caption{Mean average precision (AP) and mean precision at recall = 0.90 for each benchmark base rate.
Values are reported as mean $\pm$ error, where the error reflects the maximum absolute deviation from the 95\% bootstrap confidence interval.}
\label{tab:precision-by-model}
\end{table*}

We implemented several representative approaches for real-time phishing detecting,
including:
(1) a linear model trained on n-gram and hand-crafted features;
(2) a shallow feedforward neural network (FFN) using the same feature set;
(3) a BERT-based model with a classifier head fine-tuned on raw HTML and URLs; and,
(4) a large language model (LLM) used in a zero-shot prediction setting.
For the linear and FFN models, input features include tokenized character-level n-grams from both the URL and HTML content, as well as hand-engineered features such as the length of the URL, number of nodes in the HTML DOM tree, and other domain-specific features.
In contrast, the BERT and LLM-based approaches operate directly on raw textual inputs to learn rich representations.
Our objective is not to optimize for state-of-the-art performance, but
rather to establish baseline results on the benchmark datasets using
common solution approaches.
Model architecture and training details for each approach are described in Appendix~\ref{sec:baseline-setup}.

\textbf{Linear model}
The linear model is a simple linear classifier $f(x) = W \cdot x + b$ implemented using \texttt{LinearSVC} from the scikit-learn library \cite{scikit-learn} where parameters $W$ and $b$ are learned by minimizing the L2-regularized squared hinge loss:  
$
\mathcal{L}(W, b) = \frac{1}{N} \sum_{i=1}^{N} (\max(0, 1 - y_i (W \cdot x_i + b)))^2 + \frac{\lambda}{2} \|W\|^2_2
$.

\textbf{FFN model}
The FFN model uses the same input representation as the linear model but replaces the linear scoring function with a shallow neural network comprising of a single hidden layer with 32 neurons and ReLU activation.
Input features are binarized and subject to frequency-based feature selection to reduce dimensionality.
The model outputs a scalar probability estimate $f(x) = \sigma(W_2 \cdot \texttt{ReLU}(W_1 \cdot x + b_1) + b_2)$.
This architecture introduces a small amount of non-linearity while maintaining tractable training and inference costs.
Training is performed using binary cross-entropy loss with the Adam optimizer at a learning rate of 0.001.


\textbf{BERT-based feature learning}
In this approach, we tokenize the raw, concatenated URL $u \in \mathcal{U}$
and HTML $h \in \mathcal{H}$ and input to a BERT-based encoder model, which
generates a hidden contextualized representation,
$\mathbf{h} = \texttt{Encoder}(\texttt{tokenize}(u \oplus h)) \in \mathbb{R}^{n \times d}$,
where $n$ is the number of tokens and $d$ is encoder model dimension.
In particular, we use the \texttt{gte-large} pre-trained embedding model \cite{li2023towards},
which computes the dense embedding as
$x = \frac{1}{n}\sum_{i=1}^{n} h_i \in \mathbb{R}^d$.
We attach a linear classification head for phishing detection,
$f(x) = \sigma(W \cdot x + b) \in [0,1]$. 
We fully fine-tune the model using a binary cross-entropy loss
for two epochs on the training dataset.

\textbf{LLM-based prediction}
We prompt an LLM (Google's Gemini-2.5-Pro) with a webpage URL and its procressed HTML to produce
a rating indicating the likelihood that the page is a phishing
site. The prompt is zero-shot, and the HTML is pruned to remove
non-informative elements such as JavaScript, CSS and comments,
inspired by \cite{koide2024chatphishdetector}.  This reduces input
tokens and filters out content that is likely to be irrelevant to
phishing detection.
Further details, including the full prompt, are provided in
Appendix~\ref{sec:baseline-setup}.

\subsection{Results and Discussion}

\textbf{Performance Metrics and Evaluation Goals}
Users downstream of phishing detection systems are typically highly sensitive to
false positives -- instances where access to legitimate websites is
incorrectly blocked. This is quantified by precision, the fraction of times
users are correctly blocked. Consequently, commercial products aspire
to high values of precision while also maintaining a good
recall. Accordingly, we evaluate our models on two performance
metrics -- the average precision (AP), which approximates the area
under the precision-recall curve and the precision at a fixed recall of 0.9
($P@R=0.9$).

\textbf{Impact of Base Rate on Performance}
We evaluate the models on the test dataset (base rate; 46\%),
the full benchmark dataset (base rate: 24\%) 
and the benchmark datasets with the five base rates ranging from 0.05\%
to 5\%. Table~\ref{tab:precision-by-model} summarizes the results of
the four models across the different base rates, while Figure~\ref{fig:pr-bench}
depicts the precision-recall curves. For each rate, the solid line shows
the average precision-recall curve of all instances of benchmarks at that base rate.  
As expected, all models perform strongly on the full benchmark,
with most models achieving an AP of more than 95\%; the models achieve an even higher
AP of greater then 98\% on the test set.
(For detailed results on the test set, see Section \ref{sec:test-results} in the Appendix).
The results clearly demonstrate the dependence of model precision on
the base rate. As shown in Figure~\ref{fig:pr-bench}, the performance
of all models degrades as the base rate decreases. This behavior is
theoretically expected (see Section \ref{sec:baseRateDependence}); as the ratio of phish to benign 
decreases, maintaining high precision at the same recall requires a significantly lower
false positive rate (FPR). Further, the degradation is non-uniform
across differnt models.

\textbf{Model Comparison}
Overall, the GTE model performs the best.
In the most challenging scenario (base rate = 0.05),
the GTE model achieves a significantly higher AP than the other models. 
We believe LLM performs poorly compared to the Bert-based embedding model (GTE),
primarily since the LLM operates in a zero-shot setting while the embedding
model is fully fine-tuned on the training data set. 
More evaluation results, including precision-recall curves and
confusion matrices at multiple threshold values, are provided
in Appendix~\ref{sec:baseline-setup}.

\section{Conclusions}
\label{sec:conclusions}

We introduced \textit{PhreshPhish}, which to the best of our
knowledge, constitutes the largest, highest-quality public corpus of phishing
websites available.
Existing phishing website datasets often suffer from low quality due
to the inherent challenges of scraping phishing websites.
To address this, we implemented a robust scraping pipeline and then
systemetically post-processed the collected data to eliminate invalid
and mislabeled samples (see Algorithm~\ref{alg:cleaning} in Appendix~\ref{sec:appendix-dataset}). We also 
minimized train-test leakage (see Algorithm~\ref{alg:test-set} in Appendix~\ref{sec:appendix-bench}),
another concern with existing datasets.
Most published work on phishing detection report excellent results --
primarily due to use of unrealistically high base rates. As expected, we
found that model performance degrades sharply as the base
rate decreases to more realistic values.
To enable more realistic evaluation of phishing models, we constructed
975 benchmark datasets with real-world base rates ranging from 0.05\%
to 5\%, while also increasing task difficulty and diversity (see
Algorithm~\ref{alg:benchmark-set} in Appendix~\ref{sec:appendix-bench}).
Our evaluation of common detection approaches on the benchmark
datasets shows significant room for improvement.
Given the importance of false positives on user experience, we
adopt average precision and precision@recall=R as key performance
metrics. We make our dataset and benchmarks publicly available on
Hugging Face. The code used to scrape and process the dataset,
construct the test and benchmark datasets, and to build and evaluate
the baseline models is publicly available on Github
\url{https://github.com/phreshphish/phreshphish}.
Finally, to keep the dataset fresh, we plan to release versioned datasets
and benchmarks periodically. 


\newpage
\bibliographystyle{plain}
\bibliography{refs,phish-refs}

@article{li2023towards,
  title={Towards general text embeddings with multi-stage contrastive learning},
  author={Li, Zehan and Zhang, Xin and Zhang, Yanzhao and Long, Dingkun and Xie, Pengjun and Zhang, Meishan},
  journal={arXiv preprint arXiv:2308.03281},
  year={2023}
}

@article{aljofey2022effective,
  title={An effective detection approach for phishing websites using URL and HTML features},
  author={Aljofey, Ali and Jiang, Qingshan and Rasool, Abdur and Chen, Hui and Liu, Wenyin and Qu, Qiang and Wang, Yang},
  journal={Scientific Reports},
  volume={12},
  number={1},
  pages={8842},
  year={2022},
  publisher={Nature Publishing Group UK London}
}

@article{shahrivari2020phishing,
  title={Phishing detection using machine learning techniques},
  author={Shahrivari, Vahid and Darabi, Mohammad Mahdi and Izadi, Mohammad},
  journal={arXiv preprint arXiv:2009.11116},
  year={2020}
}

@inproceedings{abdelnabi2020visualphishnet,
  title={Visualphishnet: Zero-day phishing website detection by visual similarity},
  author={Abdelnabi, Sahar and Krombholz, Katharina and Fritz, Mario},
  booktitle={Proceedings of the 2020 ACM SIGSAC conference on computer and communications security},
  pages={1681--1698},
  year={2020}
}

@article{mohammad2015phishing,
  title={Phishing websites features},
  author={Mohammad, Rami M and Thabtah, Fadi and McCluskey, Lee},
  journal={School of Computing and Engineering, University of Huddersfield},
  volume={138},
  year={2015}
}

@inproceedings{liu2023knowledge,
  title={Knowledge expansion and counterfactual interaction for $\{$Reference-Based$\}$ phishing detection},
  author={Liu, Ruofan and Lin, Yun and Zhang, Yifan and Lee, Penn Han and Dong, Jin Song},
  booktitle={32nd USENIX Security Symposium (USENIX Security 23)},
  pages={4139--4156},
  year={2023}
}

@inproceedings{liu2024less,
  title={Less defined knowledge and more true alarms: Reference-based phishing detection without a pre-defined reference list},
  author={Liu, Ruofan and Lin, Yun and Teoh, Xiwen and Liu, Gongshen and Huang, Zhiyong and Dong, Jin Song},
  booktitle={33rd USENIX Security Symposium (USENIX Security 24)},
  pages={523--540},
  year={2024}
}

@inproceedings{afroz2011phishzoo,
  title={Phishzoo: Detecting phishing websites by looking at them},
  author={Afroz, Sadia and Greenstadt, Rachel},
  booktitle={2011 IEEE fifth international conference on semantic computing},
  pages={368--375},
  year={2011},
  organization={IEEE}
}

@inproceedings{lin2021phishpedia,
  title={Phishpedia: A hybrid deep learning based approach to visually identify phishing webpages},
  author={Lin, Yun and Liu, Ruofan and Divakaran, Dinil Mon and Ng, Jun Yang and Chan, Qing Zhou and Lu, Yiwen and Si, Yuxuan and Zhang, Fan and Dong, Jin Song},
  booktitle={30th USENIX Security Symposium (USENIX Security 21)},
  pages={3793--3810},
  year={2021}
}

@article{li2024phishintel,
  title={PhishIntel: Toward Practical Deployment of Reference-based Phishing Detection},
  author={Li, Yuexin and Tan, Hiok Kuek and Meng, Qiaoran and Lock, Mei Lin and Cao, Tri and Deng, Shumin and Oo, Nay and Lim, Hoon Wei and Hooi, Bryan},
  journal={arXiv preprint arXiv:2412.09057},
  year={2024}
}

@inproceedings{li2024knowphish,
  title={$\{$KnowPhish$\}$: Large Language Models Meet Multimodal Knowledge Graphs for Enhancing $\{$Reference-Based$\}$ Phishing Detection},
  author={Li, Yuexin and Huang, Chengyu and Deng, Shumin and Lock, Mei Lin and Cao, Tri and Oo, Nay and Lim, Hoon Wei and Hooi, Bryan},
  booktitle={33rd USENIX Security Symposium (USENIX Security 24)},
  pages={793--810},
  year={2024}
}

@article{zieni2023phishing,
  title={Phishing or not phishing? A survey on the detection of phishing websites},
  author={Zieni, Rasha and Massari, Luisa and Calzarossa, Maria Carla},
  journal={IEEE Access},
  volume={11},
  pages={18499--18519},
  year={2023},
  publisher={IEEE}
}

@article{koide2024chatphishdetector,
  title={ChatPhishDetector: Detecting Phishing Sites Using Large Language Models},
  author={Koide, Takashi and Nakano, Hiroki and Chiba, Daiki},
  journal={IEEE Access},
  year={2024},
  publisher={IEEE}
}

@article{li2024state,
  title={A State-of-the-art Review on Phishing Website Detection Techniques},
  author={Li, Wenhao and Manickam, Selvakumar and Chong, Yung-Wey and Leng, Weilan and Nanda, Priyadarsi},
  journal={IEEE Access},
  year={2024},
  publisher={IEEE}
}

@misc{cyberkillchain,
    author = "{Wikipedia contributors}",
    title = "Cyber kill chain --- {Wikipedia}{,} The Free Encyclopedia",
    year = "2024",
    howpublished = "\url{https://en.wikipedia.org/w/index.php?title=Cyber_kill_chain&oldid=1256319774}",
    note = "[Online; accessed 6-April-2025]"
  }

@inproceedings{bijmans2021catching,
  title={Catching phishers by their bait: Investigating the dutch phishing landscape through phishing kit detection},
  author={Bijmans, Hugo and Booij, Tim and Schwedersky, Anneke and Nedgabat, Aria and van Wegberg, Rolf},
  booktitle={30th USENIX security symposium (USENIX security 21)},
  pages={3757--3774},
  year={2021}
}

@article{cova2008there,
  title={There Is No Free Phish: An Analysis of" Free" and Live Phishing Kits.},
  author={Cova, Marco and Kruegel, Christopher and Vigna, Giovanni},
  journal={WOOT},
  volume={8},
  pages={1--8},
  year={2008}
}

@article{nagunwa2022machine,
  title={A machine learning approach for detecting fast flux phishing hostnames},
  author={Nagunwa, Thomas and Kearney, Paul and Fouad, Shereen},
  journal={Journal of information security and applications},
  volume={65},
  pages={103125},
  year={2022},
  publisher={Elsevier}
}

@inproceedings{penedo2024fineweb,
 author = {Penedo, Guilherme and Kydl\'{\i}\v{c}ek, Hynek and allal, Loubna Ben and Lozhkov, Anton and Mitchell, Margaret and Raffel, Colin and Von Werra, Leandro and Wolf, Thomas},
 booktitle = {Advances in Neural Information Processing Systems},
 editor = {A. Globerson and L. Mackey and D. Belgrave and A. Fan and U. Paquet and J. Tomczak and C. Zhang},
 pages = {30811--30849},
 publisher = {Curran Associates, Inc.},
 title = {The FineWeb Datasets: Decanting the Web for the Finest Text Data at Scale},
 url = {https://proceedings.neurips.cc/paper_files/paper/2024/file/370df50ccfdf8bde18f8f9c2d9151bda-Paper-Datasets_and_Benchmarks_Track.pdf},
 volume = {37},
 year = {2024}
}

@misc{kaggle-huntingdata11,
  author       = {Phishing Website HTML Classification},
  howpublished = {\url{https://www.kaggle.com/datasets/huntingdata11/phishing-website-html-classification}},
  note         = {Accessed: 2025-04-20}
}

@misc{kaggle-haozhang1579,
  author       = {Crawling2024(Phishing Websites Dataset)},
  howpublished = {\url{https://www.kaggle.com/datasets/haozhang1579/crawling2024}},
  note         = {Accessed: 2025-04-20}
}

@misc{kaggle-jackcavar,
  author       = {Malicious and Benign Website dataset},
  howpublished = {\url{ https://www.kaggle.com/datasets/jackcavar/malicious-and-benign-website-dataset}},
  note         = {Accessed: 2025-04-20}
}

@misc{kaggle-aljofey,
  author       = {Phishing Data},
  howpublished = {\url{https://www.kaggle.com/datasets/aljofey/phishing-data}},
  note         = {Accessed: 2025-04-20}
}

@misc{huggingface-ealvaradob,
  author       = {Phishing Dataset},
  howpublished = {\url{https://huggingface.co/datasets/ealvaradob/phishing-dataset}},
  note         = {Accessed: 2025-04-20}
}

@misc{huggingface-huynq3Cyradar,
  author       = {Phishing Detection Dataset},
  howpublished = {\url{https://huggingface.co/datasets/huynq3Cyradar/Phishing_Detection_Dataset}},
  note         = {Accessed: 2025-04-20}
}

@misc{apwg,
  author       = {Anti-Phishing Working Group (APWG)},
  howpublished = {\url{https://apwg.org/about-us/}},
  note         = {Accessed: 2025-04-20}
}

@misc{apwg-report-2025q1,
    author = {Anti-Phishing Working Group},
    title = "Phishing Attack Trends Report -- 2025 Q1",
    year = "2025",
    howpublished = {\url{https://apwg.org/trendsreports/}},
    note = "[Online; accessed 2026-02-07]"
  }

@misc{fbi-ic3report-2024,
    author = "{FBI - Internet Crime Complaint Center (IC3)}",
    title = "2024 {IC3} Annual Report",
    year = "2024",
    howpublished = {\url{https://www.ic3.gov/AnnualReport/Reports}},
    note = "[Online; accessed 2026-02-07]"
  }

@misc{dark-web-2018,
    author = "Simon Migliano",
    title = "The Dark Web is Democratizing Cybercrime",
    year = "2018",
    month = "August",
    howpublished = {\url{https://hackernoon.com/the-dark-web-is-democratizing-cybercrime75e951e2454}},
    note = "[Online; accessed 2025-04-20]"
  }

@article{kaufman2012leakage,
  title={Leakage in data mining: Formulation, detection, and avoidance},
  author={Kaufman, Shachar and Rosset, Saharon and Perlich, Claudia and Stitelman, Ori},
  journal={ACM Transactions on Knowledge Discovery from Data (TKDD)},
  volume={6},
  number={4},
  pages={1--21},
  year={2012},
  publisher={ACM New York, NY, USA}
}

@article{kapoor2022leakage,
  title={Leakage and the reproducibility crisis in ML-based science},
  author={Kapoor, Sayash and Narayanan, Arvind},
  journal={arXiv preprint arXiv:2207.07048},
  year={2022}
}

@article{oner2020training,
  title={Training machine learning models on patient level data segregation is crucial in practical clinical applications},
  author={Oner, Mustafa Umit and Cheng, Yi-Chih and Lee, Hwee Kuan and Sung, Wing-Kin},
  journal={medRxiv},
  pages={2020--04},
  year={2020},
  publisher={Cold Spring Harbor Laboratory Press}
}

@article{ma2011learning,
  title={Learning to detect malicious urls},
  author={Ma, Justin and Saul, Lawrence K and Savage, Stefan and Voelker, Geoffrey M},
  journal={ACM Transactions on Intelligent Systems and Technology (TIST)},
  volume={2},
  number={3},
  pages={1--24},
  year={2011},
  publisher={ACM New York, NY, USA}
}

@inproceedings{ma2009beyond,
  title={Beyond blacklists: learning to detect malicious web sites from suspicious URLs},
  author={Ma, Justin and Saul, Lawrence K and Savage, Stefan and Voelker, Geoffrey M},
  booktitle={Proceedings of the 15th ACM SIGKDD international conference on Knowledge discovery and data mining},
  pages={1245--1254},
  year={2009}
}

@article{scikit-learn,
  title={Scikit-learn: Machine Learning in {P}ython},
  author={Pedregosa, F. and Varoquaux, G. and Gramfort, A. and Michel, V.
          and Thirion, B. and Grisel, O. and Blondel, M. and Prettenhofer, P.
          and Weiss, R. and Dubourg, V. and Vanderplas, J. and Passos, A. and
          Cournapeau, D. and Brucher, M. and Perrot, M. and Duchesnay, E.},
  journal={Journal of Machine Learning Research},
  volume={12},
  pages={2825--2830},
  year={2011}
}

@article{andoni2008near,
  title={Near-optimal hashing algorithms for approximate nearest neighbor in high dimensions},
  author={Andoni, Alexandr and Indyk, Piotr},
  journal={Communications of the ACM},
  volume={51},
  number={1},
  pages={117--122},
  year={2008},
  publisher={ACM New York, NY, USA}
}

@article{das2019sok,
  title={SoK: a comprehensive reexamination of phishing research from the security perspective},
  author={Das, Avisha and Baki, Shahryar and El Aassal, Ayman and Verma, Rakesh and Dunbar, Arthur},
  journal={IEEE Communications Surveys \& Tutorials},
  volume={22},
  number={1},
  pages={671--708},
  year={2019},
  publisher={IEEE}
}

@inproceedings{whittaker2010large,
  title={Large-Scale Automatic Classification of Phishing Pages.},
  author={Whittaker, Colin and Ryner, Brian and Nazif, Marria},
  booktitle={Ndss},
  volume={10},
  pages={2010},
  year={2010}
}

@article{tang2021survey,
  title={A survey of machine learning-based solutions for phishing website detection},
  author={Tang, Lizhen and Mahmoud, Qusay H},
  journal={Machine Learning and Knowledge Extraction},
  volume={3},
  number={3},
  pages={672--694},
  year={2021},
  publisher={MDPI}
}

@inproceedings{cao2025phishagent,
  title={Phishagent: a robust multimodal agent for phishing webpage detection},
  author={Cao, Tri and Huang, Chengyu and Li, Yuexin and Huilin, Wang and He, Amy and Oo, Nay and Hooi, Bryan},
  booktitle={Proceedings of the AAAI Conference on Artificial Intelligence},
  volume={39},
  pages={27869--27877},
  year={2025}
}

@article{zhang2025benchmarking,
  title={Benchmarking and Evaluating Large Language Models in Phishing Detection for Small and Midsize Enterprises: A Comprehensive Analysis},
  author={Zhang, Jun and Wu, Peiqiao and London, Jeffrey and Tenney, Dan},
  journal={IEEE Access},
  year={2025},
  publisher={IEEE}
}

@article{el2020depth,
  title={An in-depth benchmarking and evaluation of phishing detection research for security needs},
  author={El Aassal, Ayman and Baki, Shahryar and Das, Avisha and Verma, Rakesh M},
  journal={Ieee Access},
  volume={8},
  pages={22170--22192},
  year={2020},
  publisher={IEEE}
}

@article{patterson2012better,
  title={For better or worse, benchmarks shape a field},
  author={Patterson, David},
  journal={Communications of the ACM},
  volume={55},
  year={2012}
}

@misc{selenium,
  author       = {Selenium Grid},
  howpublished = {\url{https://www.selenium.dev/documentation/grid/}},
  note         = {Accessed: 2025-03-27}
}

@misc{phishtank,
  author       = {PhishTank},
  howpublished = {\url{https://phishtank.org}},
  note         = {Accessed: 2025-03-28}
}

@misc{apwgecrimex,
  author       = {eCrime eXchange},
  howpublished = {\url{https://apwg.org}},
  note         = {Accessed: 2025-03-28}
}

@misc{netcraft,
  author       = {Netcraft},
  howpublished = {\url{https://www.netcraft.com}},
  note         = {Accessed: 2025-03-28}
}

@misc{nutch,
  author       = {Apache Nutch},
  howpublished = {\url{https://nutch.apache.org/}},
  note         = {Accessed: 2025-04-16}
}

@misc{commoncrawl,
  author       = {Common Crawl},
  howpublished = {\url{https://commoncrawl.org}},
  note         = {Accessed: 2025-04-17}
}

@article{raffel2020exploring,
  title   = {Exploring the limits of transfer learning with a unified text-to-text transformer},
  author  = {Raffel, Colin and Shazeer, Noam and Roberts, Adam and Lee, Katherine and Narang, Sharan and Matena, Michael and Zhou, Yanqi and Li, Wei and Liu, Peter J},
  journal = {Journal of machine learning research},
  volume  = {21},
  number  = {140},
  pages   = {1--67},
  year    = {2020}
}

@misc{react,
  author       = {React},
  howpublished = {\url{https://react.dev}},
  note         = {Accessed: 2025-04-19}
}

@misc{angular,
  author       = {Angular},
  howpublished = {\url{https://angular.dev}},
  note         = {Accessed: 2025-04-19}
}

@inproceedings{9519414,
  author    = {Zhang, Penghui and Oest, Adam and Cho, Haehyun and Sun, Zhibo and Johnson, RC and Wardman, Brad and Sarker, Shaown and Kapravelos, Alexandros and Bao, Tiffany and Wang, Ruoyu and Shoshitaishvili, Yan and Doupé, Adam and Ahn, Gail-Joon},
  booktitle = {2021 IEEE Symposium on Security and Privacy (SP)},
  title     = {CrawlPhish: Large-scale Analysis of Client-side Cloaking Techniques in Phishing},
  year      = {2021},
  volume    = {},
  number    = {},
  pages     = {1109-1124},
  doi       = {10.1109/SP40001.2021.00021}
}

@article{tian2022comprehensive,
  title={A comprehensive survey on poisoning attacks and countermeasures in machine learning},
  author={Tian, Zhiyi and Cui, Lei and Liang, Jie and Yu, Shui},
  journal={ACM Computing Surveys},
  volume={55},
  number={8},
  pages={1--35},
  year={2022},
  publisher={ACM New York, NY}
}

@misc{graphql,
  author       = {GraphQL},
  howpublished = {\url{https://graphql.org}},
  note         = {Accessed: 2025-04-19}
}

@article{xie2025scalable,
  title={A scalable phishing website detection model based on dual-branch TCN and mask attention},
  author={Xie, Lixia and Zhang, Hao and Yang, Hongyu and Hu, Ze and Cheng, Xiang},
  journal={Computer Networks},
  volume={263},
  pages={111230},
  year={2025},
  publisher={Elsevier}
}

@misc{datacommons,
  author       = {Data Commons},
  howpublished = {\url{https://datacommons.org}},
  note         = {Accessed: 2025-05-19}
}

@article{aleroud2017phishing,
  title={Phishing environments, techniques, and countermeasures: A survey},
  author={Aleroud, Ahmed and Zhou, Lina},
  journal={Computers \& Security},
  volume={68},
  pages={160--196},
  year={2017},
  publisher={Elsevier}
}

@techreport{enisa2025threatLandscape,
  author       = {{European Union Agency for Cybersecurity (ENISA)}},
  title        = {ENISA Threat Landscape 2025},
  institution  = {ENISA},
  year         = {2025},
  url          = {https://www.enisa.europa.eu/publications/enisa-threat-landscape-2025}
}

\newpage

\appendix

\section{Our Dataset}
\label{sec:appendix-dataset}

\begin{figure*}[ht]
    \centering
    \includegraphics[width=\linewidth]{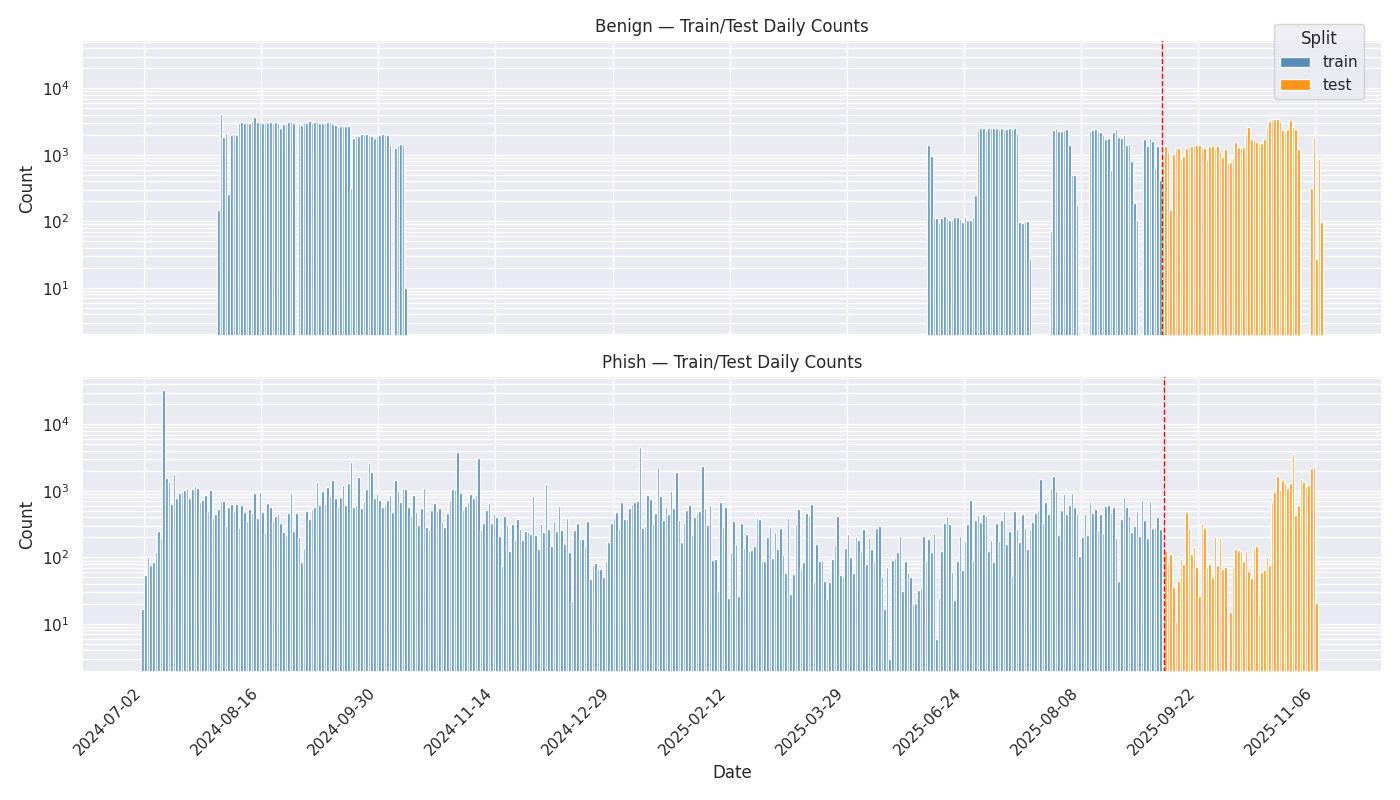}
    \caption{The train and test datasets are temporally disjoint for each class.}
    \label{fig:dataset-temporal}
\end{figure*}

Our data collection system runs continuously, collecting URLs from multiple feeds,
and scraping the corresponding HTML content using a real browser to ensure high fidelity.
Figure~\ref{fig:dataset-temporal} provides a temporal view of the dataset
after the cleaning process, aggregated at a daily granularity.
The volume of phishing data collected each day depended primarily
on the number of phishing pages reported by the feeds, as well as
occasional local infrastructure issues. In contrast, the benign
data collection shows greater variability, partly because benign
URLs are more readily available -- allowing a large number to be collected in
a single day -- and also because benign pages tend to be less dynamic and
remain fresh for longer periods than phishing pages. This accounts
for the periods during which no benign data was collected. 

\subsection{Cleaning details}
\begin{figure}[h]
    \
    \begin{subfigure}[t]{0.48\columnwidth}
        \includegraphics[width=\linewidth]{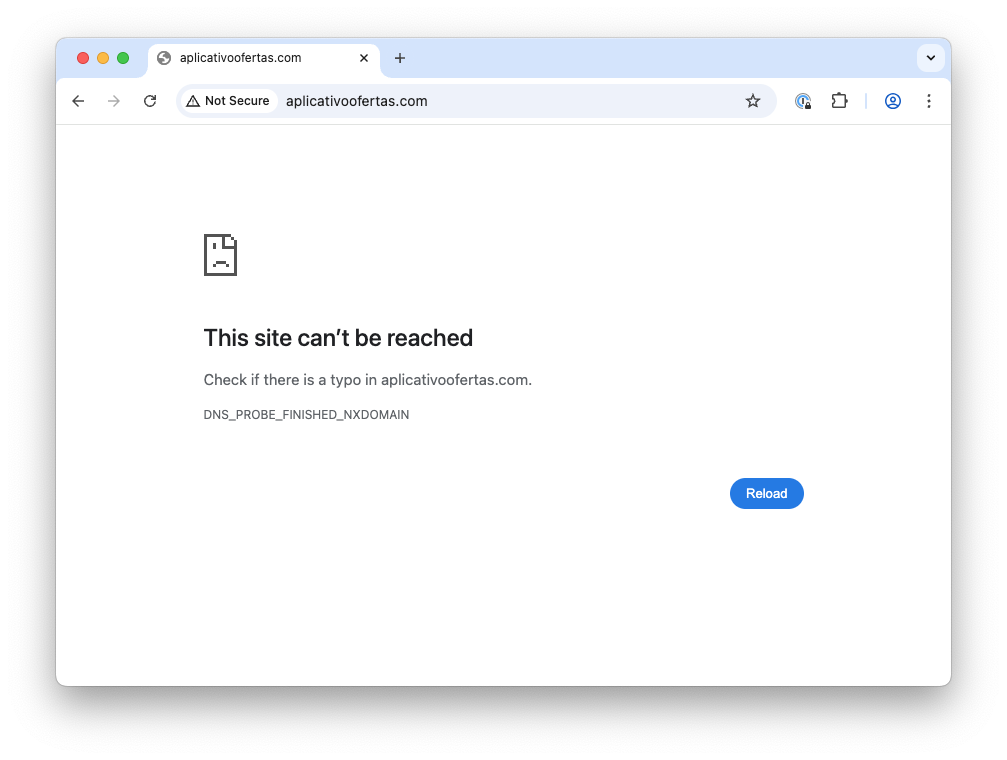}
        \caption{\label{fig:cannot-resolve}}
    \end{subfigure}
    \begin{subfigure}[t]{0.48\columnwidth}
        \includegraphics[width=\linewidth]{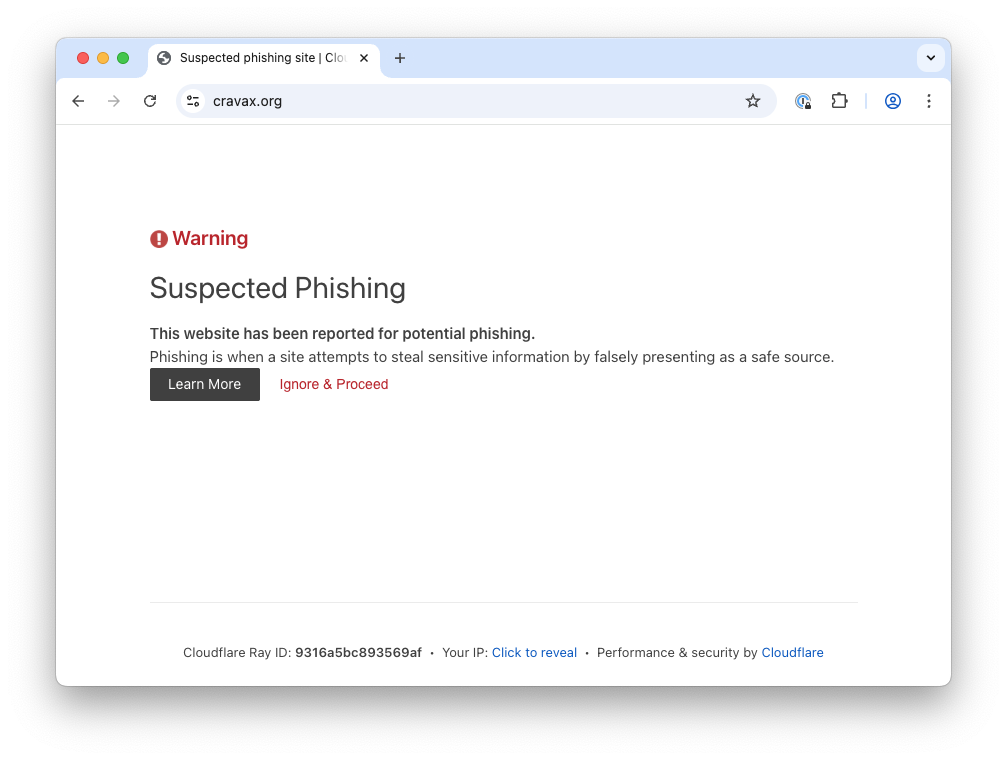}
        \caption{\label{fig:cloudflare-phishing}}
    \end{subfigure}
    \vspace{-1em}
    \begin{subfigure}[t]{0.48\columnwidth}
        \includegraphics[width=\linewidth]{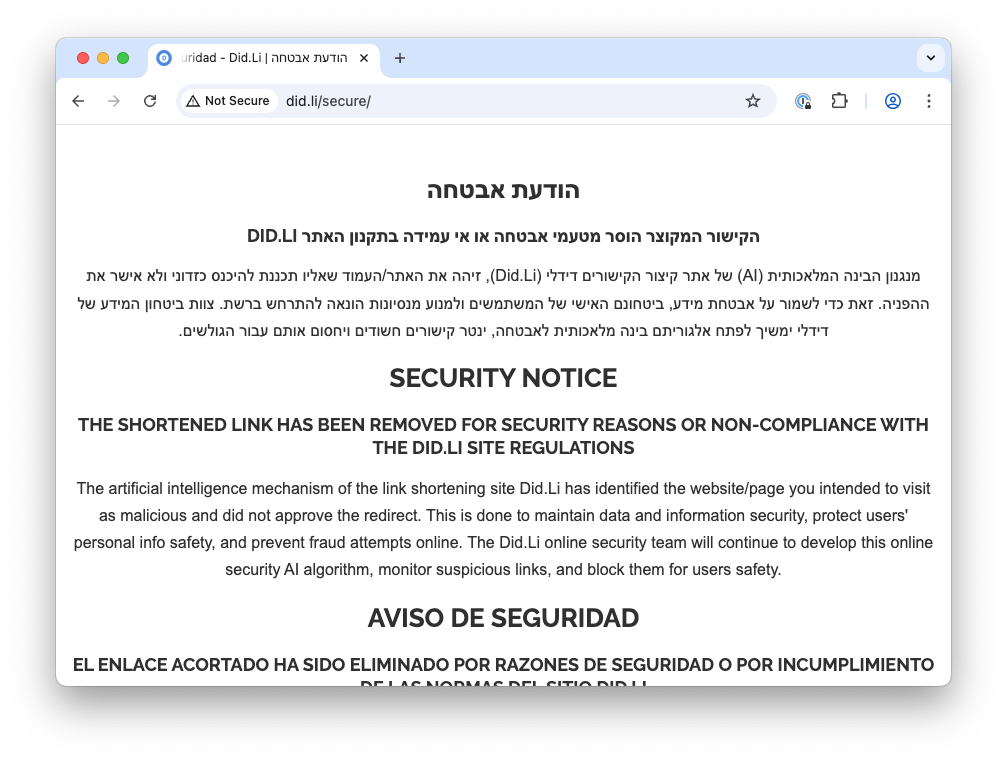}
        \caption{\label{fig:shortener-removed}}
    \end{subfigure}
    \begin{subfigure}[t]{0.48\columnwidth}
        \includegraphics[width=\linewidth]{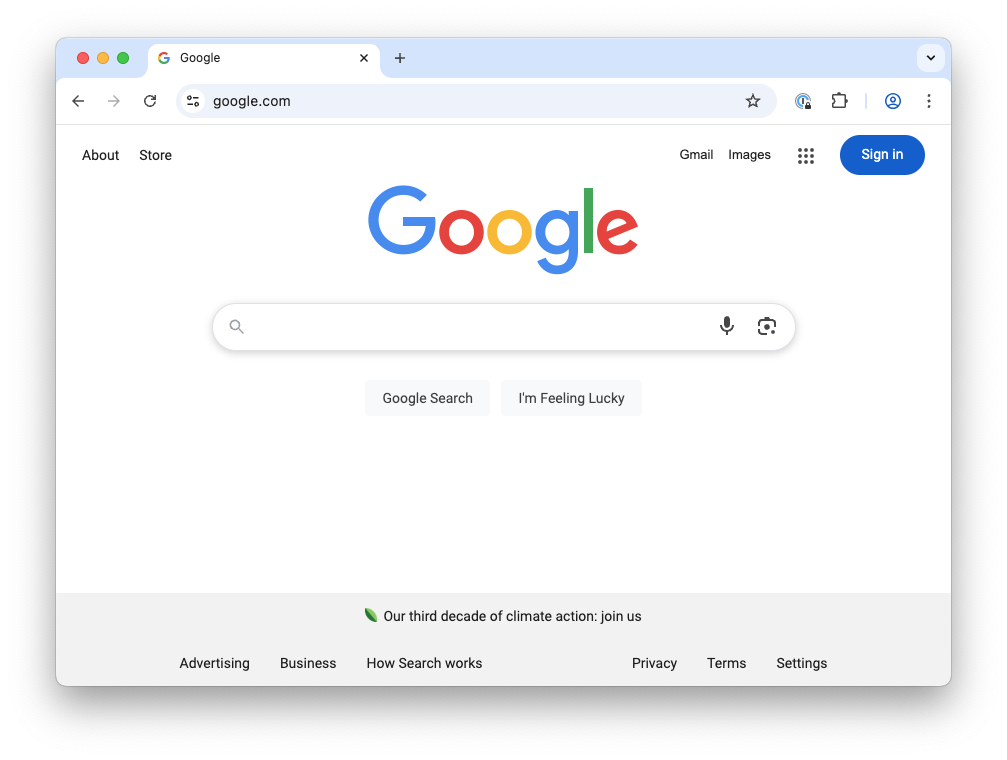}
        \caption{\label{fig:redirect}}
    \end{subfigure}
    \caption{Multiple failure modes are frequently encountered when scraping phishing pages.
             (\subref{fig:cannot-resolve}) DNS resolution issues can prevent the page from being scraped.
             (\subref{fig:cloudflare-phishing}) Security takedown notices can prevent the page from being scraped.
             (\subref{fig:shortener-removed}) URL shortening services will sometimes block access to the intended page.
             (\subref{fig:redirect}) Redirects to the legitimate target brand page can occur resulting in mislabeled data. 
             }
    \label{fig:failure-modes}
\end{figure}

Cleaning consists of two stages: (1) a fully automated stage that removes pages with invalid HTML content and other obvious artifacts such as bad titles as shown in Figure~\ref{tab:bad-titles}; and (2) a human-in-the-loop stage that identifies and removes pages likely to correspond to failure modes.
While our scraping system was designed to mitigate many of the challenges associated with phishing data collection described in Section~\ref{sec:challenges}, scrape failures are still common and must be removed.
Representative examples of common failure modes are shown in Figure~\ref{fig:failure-modes}.

\begin{table}[h]
  \centering
  \small
  \caption{Keywords used for filtering bad page titles. These keywords are used to automatically identify and remove low-quality samples from the dataset.}
  \label{tab:bad-titles}
  \begin{tabular}{p{0.40\columnwidth} p{0.50\columnwidth}}
    \toprule
    \textbf{Keywords} & \textbf{Example title}\\
    \midrule
    "400" & "400 Bad Request" \\
    "403" & "403 Forbidden" \\
    "404" & "404 Not Found" \\
    "410" & "410 Gone" \\
    "found" & "Page not found!" \\
    "encontrada" & "Página não encontrada" \\
    "forbidden" & "Access Forbidden" \\
    "error" & "Sorry, an error occurred" \\
    "suspended" & "Account Suspended" \\
    "bad request" & "Bad Request" \\
    "cloudflare" & "Cloudflare Anti-Bot" \\
    "just a moment" & "Just a moment..." \\
    "warning! | there might be a problem" & "Warning! | There might be a problem with the requested link" \\
    "url shortener, branded short" & "URL Shortener, Branded Short Links \& Analytics | TinyURL" \\
    "denied" & "Access Denied" \\
    \bottomrule
  \end{tabular}
\end{table}

During the manual inspection stage, pages are first vectorized using TF-IDF features (line 4).
These vectors are then projected into binary hash bins using random projection to form coarse similarity groups (lines 5 and 9).
Pages are also grouped by their HTML \texttt{<title>} tags (lines 7 and 10), which is helpful for identifying pages that are likely to be derived from the same template or boilerplate code.
Within each group, a single representative (prototype) page is selected and presented for manual inspection (lines 18 and 19).
The human annotator is then asked to determine whether the page reflects a failure mode or otherwise invalid content (line 20).
If the page represents a scrape failure, it and all of its nearest neighbors (under cosine similarity in the TF-IDF space) are removed from the dataset (line 21).
Groups are prioritized by size, ensuring that the cleaning budget is spent efficiently (lines 16 and 17).

\begin{algorithm}[h]
    \caption{Cleaning process}
    \label{alg:cleaning}
    \begin{algorithmic}[1]
        \State \textbf{Input:} Dataset $\mathcal{D}$, budget $b$
        \State \textbf{Output:} Cleaned dataset $\mathcal{D}' \subseteq \mathcal{D}$
        \State
        \State $X \gets \texttt{TFIDF}(\mathcal{D})$ \Comment{create TF-IDF features}
        \State $B \gets \big(X \cdot \mathcal{N}(0, 1)\big) \geq 0$ \Comment{bin via random projection}
        \State
        \State $T \gets \{\texttt{GetTitle}(x) \mid x \in \mathcal{D}\}$
        \State
        \State $G_B \gets \texttt{GroupBy}(B, X)$ \Comment{group by LSH bins}
        \State $G_T \gets \texttt{GroupBy}(T, X)$ \Comment{group by titles}
        \State
        \State \textit{// iterate through groups, keeping track of which pages to remove}
        \State $R \gets \emptyset$ \Comment{to remove}
        \For {$G \in \{G_B, G_T\}$}
            \State $i \gets 0$ \Comment{stop when budget exhausted}
            \While{$i < b$}
                \For{$g \in \texttt{Sorted}(G, \texttt{key}=|g|, \texttt{descending=True})$}
                    \State $p \gets \texttt{SelectPrototype}(g)$
                    \State $\texttt{doRemove} \gets \texttt{ManualInspect}(p)$
                    \If{$\texttt{doRemove}$}
                        \State $R \gets R \cup \texttt{GetNearestNeighbors}(p)$
                    \EndIf
                    \State $i \gets i + 1$
                \EndFor
            \EndWhile
        \EndFor
        \State \Return $\mathcal{D}' \gets X \setminus R$
    \end{algorithmic}
\end{algorithm}

\subsection{Handling PII}
We focused our PII removal efforts on the benign class only as this is the class that is most likely to contain sensitive information.
In fact, because phishing pages are often designed to look like legitimate pages, they often contain data such as phone numbers and email addresses which may be useful signals for phishing detectors.

Query parameters in the URL are often used to pass sensitive information such as user IDs, session tokens, and other PII to the server and are therefore removed from the URL.
We also removed data points with URLs that may contain customer information.
This is done by first transforming the URL into a string of words by replacing special characters (e.g. \texttt{/} and \texttt{-}) with spaces and then using a DistilBERT-based NER model \footnote{\url{https://huggingface.co/dslim/bert-base-NER}} to identify any URLs that contain a person's name.
We then used \texttt{SmolLM2-1.7B-Instruct} \footnote{\url{https://huggingface.co/HuggingFaceTB/SmolLM2-1.7B-Instruct}} to determine whether the name identified by the NER model is a well-known or otherwise public figure or not.
If the name is not a well-known or public figure, the URL and corresponding page is removed from the dataset.

We did not attempt to remove other PII such as email addresses or phone numbers from the benign class as this would require differentiating between true PII and data that is not PII but looks like it (e.g. a customer support phone number or email address).
Websites that tend to contain true PII (e.g. social media sites) contain such data behind authentication walls.
Because our scraping process does not attempt to log in to sites, we do not expect truly sensitive data to be present in the dataset.

\subsection{Risks and Limitations}
We found that additional kinds of data including screenshots and raw network traffic (to detect things like redirects, for example) could be useful for both cleaning as well as for training other kinds of models (e.g. reference-based image models) and hope to include these features in future versions of the dataset.
Because there are varying degrees of sophistication in phishing campaigns, there is a risk that the phishing pages we were able to scrape are systemically easier to scrape (and therefore to detect) than those that are not included in the dataset.

Finally, while our primary intent is to support research in phishing detection and web security, we acknowledge the dual-use risk of releasing this dataset.
Malicious actors could theoretically use the dataset to refine and improve their tactics.
However, we believe that openness and transparency is ultimately more beneficial to the community than secrecy.
By making the dataset and benchmarks available, we aim to foster the development of stronger defenses against phishing attacks.

\begin{figure*}
  \centering

  \setlength{\abovecaptionskip}{2pt}
  \setlength{\belowcaptionskip}{0pt}

  \begin{minipage}{\textwidth}
    \centering
    \captionsetup{type=table}
    \caption{Summary statistics for selected URL features.}
    \resizebox{\textwidth}{!}{%
      \begin{tabular}{llrrrrrrrrrr}
        \toprule
        \textbf{Feature} & \textbf{Description} 
        & \multicolumn{2}{c}{\textbf{Min}} 
        & \multicolumn{2}{c}{\textbf{25\%}} 
        & \multicolumn{2}{c}{\textbf{50\%}} 
        & \multicolumn{2}{c}{\textbf{75\%}} 
        & \multicolumn{2}{c}{\textbf{Max}} \\
                         &                             & Phish & Benign & Phish & Benign & Phish & Benign & Phish & Benign & Phish         & Benign                   \\
        URL length       & Num characters in the URL   & 12    & 5      & 31    & 42     & 39    & 55     & 58    & 74     & 25,523        & 1,414                    \\
        Domain length    & Num characters in domain    & 4     & 4      & 10    & 10     & 11    & 13     & 15    & 16     & 64            & 46                       \\
        Path length      & Num characters in the path  & 0     & 0      & 1     & 17     & 1     & 29     & 14    & 48     & 6,597         & 1,414                      \\
        Subdomain length & Num characters in subdomain & 0     & 0      & 3     & 3      & 9     & 3      & 17    & 3      & 92            & 81                       \\
        \bottomrule
      \end{tabular}
    }
    \label{tab:url-characteristics}
  \end{minipage}

  \vspace{6pt}

  \begin{minipage}{\textwidth}
    \centering
    \includegraphics[width=\textwidth]{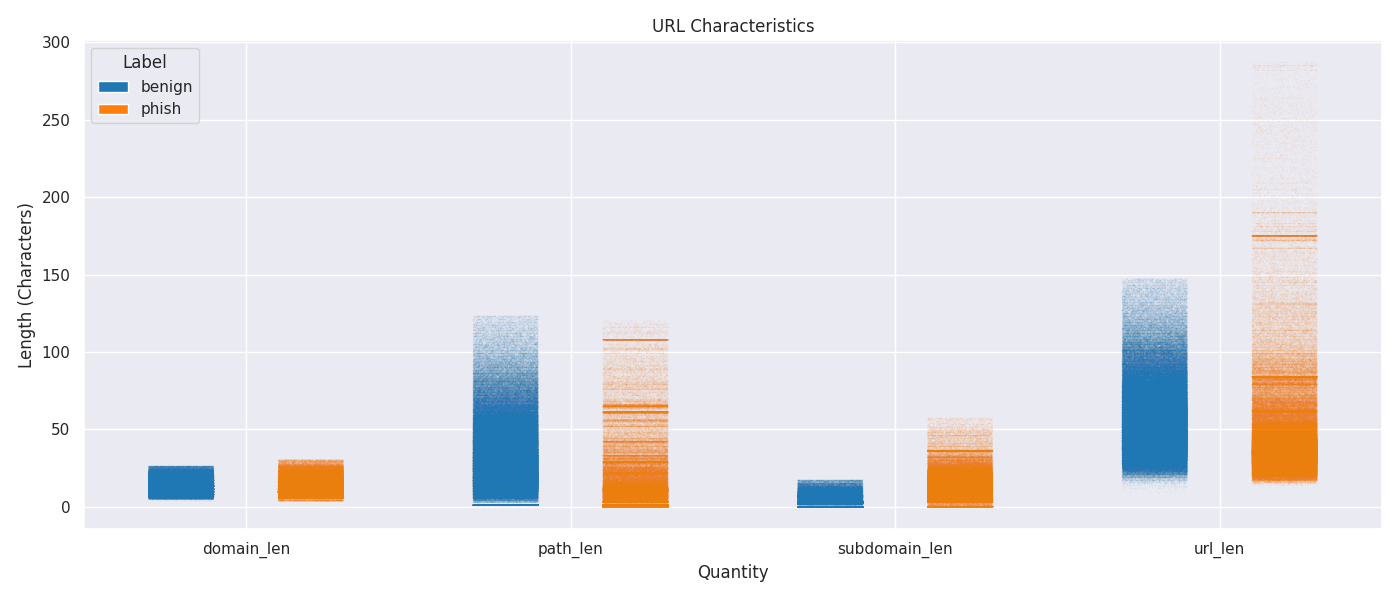}
    \captionsetup{type=figure}
    \caption{Distribution of URL characteristics with 99th percentile outliers removed.}
    \label{fig:block-1}
  \end{minipage}

  \vspace{6pt}

  \begin{minipage}{\textwidth}
    \centering
    \captionsetup{type=table}
    \caption{Summary statistics for selected HTML features.}
    \resizebox{\textwidth}{!}{%
      \begin{tabular}{llrrrrrrrrrr}
        \toprule
        \textbf{Feature} & \textbf{Description} 
        & \multicolumn{2}{c}{\textbf{Min}} 
        & \multicolumn{2}{c}{\textbf{25\%}} 
        & \multicolumn{2}{c}{\textbf{50\%}} 
        & \multicolumn{2}{c}{\textbf{75\%}} 
        & \multicolumn{2}{c}{\textbf{Max}} \\
                          &                                                       & Phish & Benign & Phish & Benign & Phish  & Benign & Phish & Benign  & Phish      & Benign     \\
        HTML length        & \textless html\textgreater{} tag lengths             & 28    & 23     & 8,271 & 90,570 & 34,672 & 249,954 & 117,473 & 545,994 & 76,259,693 & 45,877,378 \\
        Head tag length   & Sum of all \textless head\textgreater{} tag lengths   & 0     & 0      & 1,323 & 11,168 & 5,375  & 49,107  & 21,087  & 150,520 & 16,327,295 & 32,183,109 \\
        Image tag length  & Sum of all \textless img\textgreater{} tag lengths    & 0     & 0      & 0     & 895    & 397    & 4,070   & 3,154   & 12,673  & 13,288,010 & 44,236,577 \\
        Script tag length & Sum of all \textless script\textgreater{} tag lengths & 0     & 0      & 438   & 8,573  & 3,397  & 28,939  & 24,532  & 115,009 & 16,327,282 & 21,785,672 \\
        \bottomrule
      \end{tabular}
    }
    \label{tab:html-characteristics}
  \end{minipage}

  \vspace{6pt}

  \begin{minipage}{\textwidth}
    \centering
    \includegraphics[width=\textwidth]{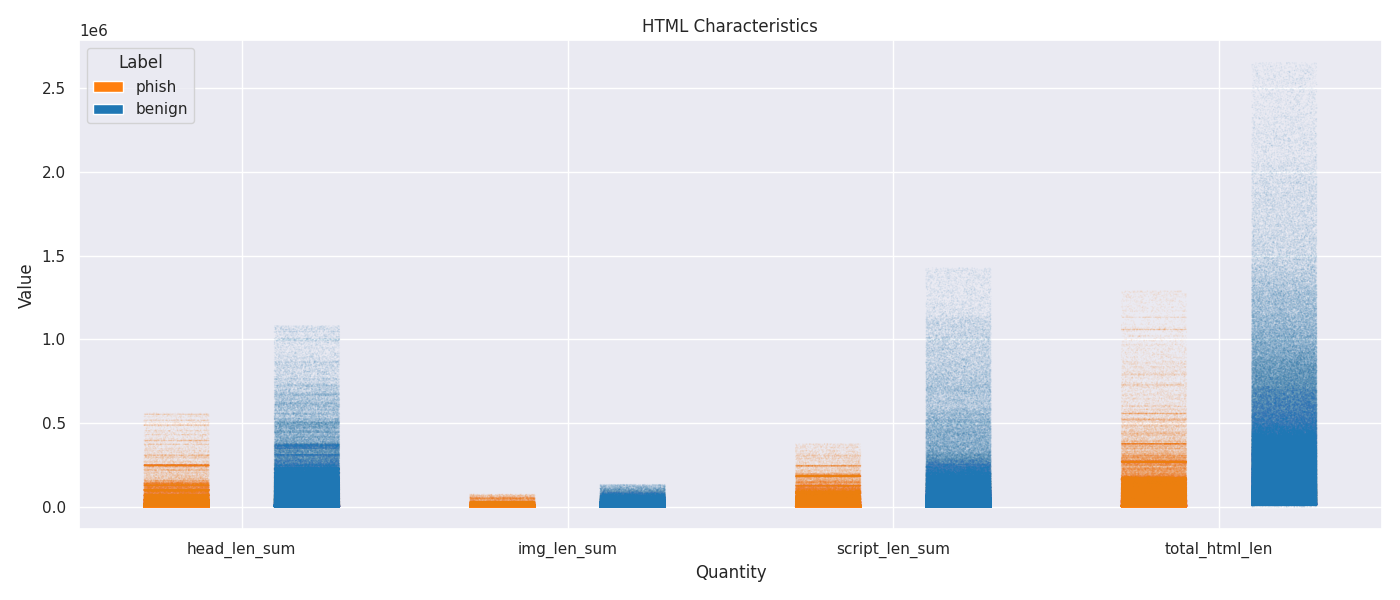}
    \captionsetup{type=figure}
    \caption{Distribution of HTML characteristics with 99th percentile outliers removed.}
    \label{fig:block-2}
  \end{minipage}

\end{figure*}

\begin{figure*}
  \centering
  \begin{subfigure}[t]{0.32\textwidth}
    \centering
    \includegraphics[width=\linewidth]{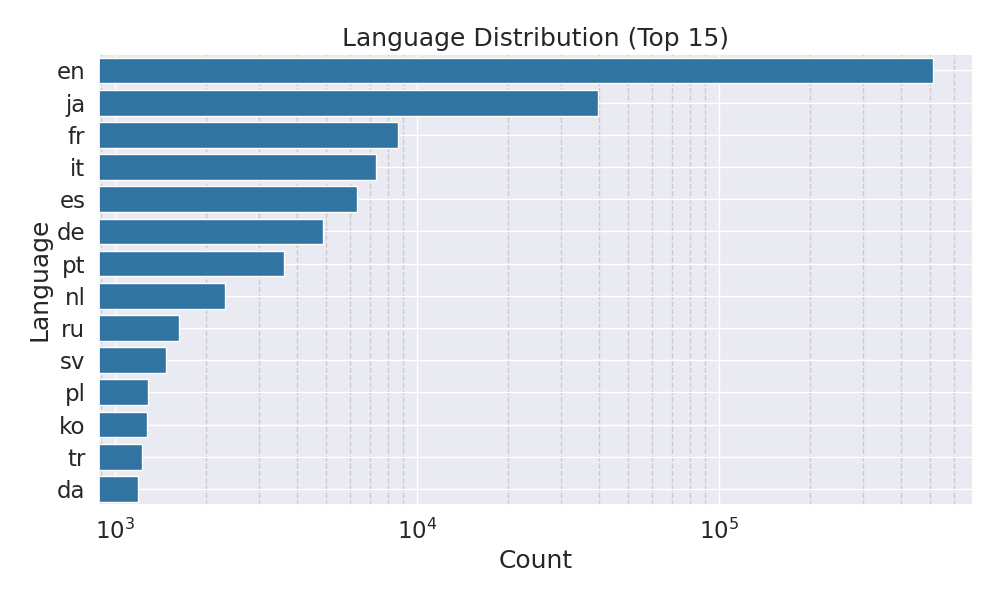}
    \caption{}
    \label{fig:lang_distribution}
  \end{subfigure}
  \hfill
  \begin{subfigure}[t]{0.32\textwidth}
    \centering
    \includegraphics[width=\linewidth]{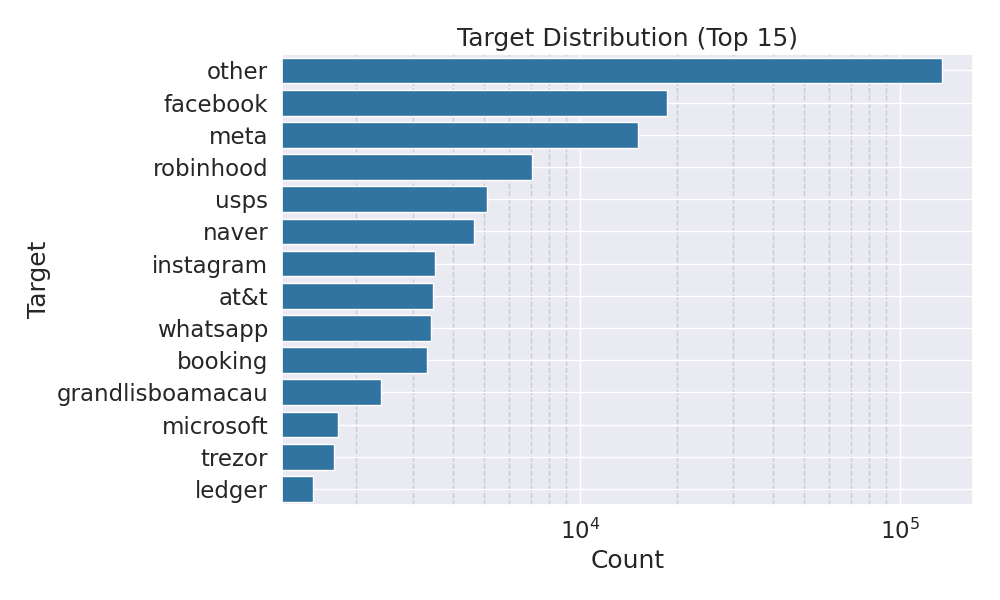}
    \caption{}
    \label{fig:top_targets}
  \end{subfigure}
  \hfill
  \begin{subfigure}[t]{0.32\textwidth}
    \centering
    \includegraphics[width=\linewidth]{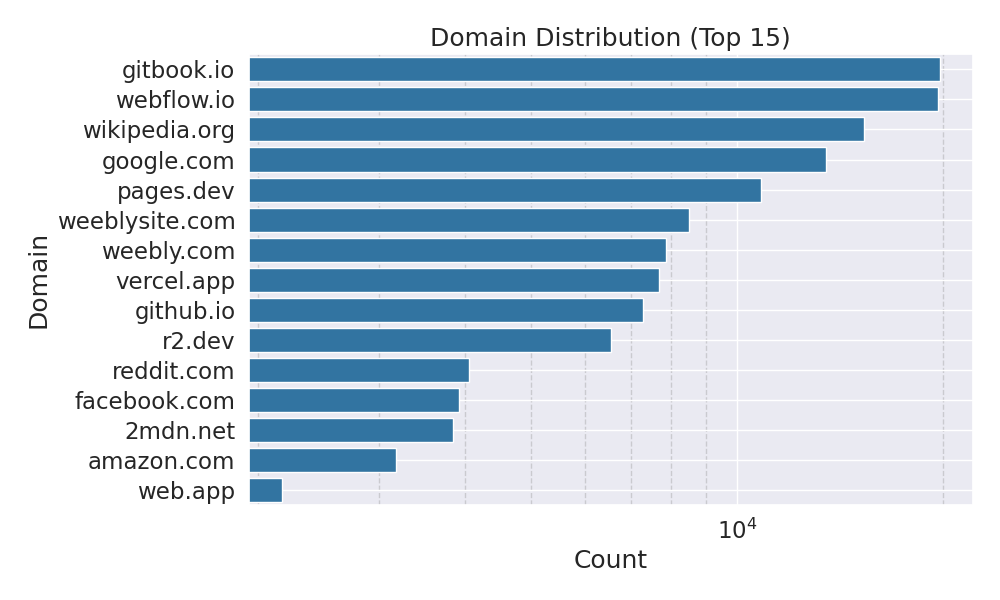}
    \caption{}
    \label{fig:domain_distribution}
  \end{subfigure}
  \caption{Dataset statistics: (a) top languages as determined by the \texttt{langdetect} library, (b) top phishing targets, and (c) top domains (both phishing and benign).}
  \label{fig:eda_overview}
\end{figure*}
\section{Test and Benchmark Sets}
\label{sec:appendix-bench}

Algorithm~\ref{alg:test-set} outlines the main procedure used to split
the dataset into training and test sets.  A test split fraction,
$p = 0.3$ was used.  Lines \texttt{5--7} create a temporal split,
performed in a stratified manner to preserve class proportions across
both sets.

Lines \texttt{10--19} minimize data leakage from training set to the test set by
iterating over  the test dataset, searching for similar instances in the
training set using LSH, and removing test points with similarity
above a threshold.

Algorithm~\ref{alg:benchmark-set} outlines the procedure for creating the
benchmark datasets from the test set.

Lines \texttt{5--6} describe the process of enhancing test dataset
diversity.  The benign part of the test set is augmented with
legitimate websites corresponding to the top target phishing brands
(determined from the training phish set).  For the phishing test data,
pairwise comparisons are performed, and one sample from each pair of
highly similar instances is removed. This pruning step eliminates
highly similar samples, increasing the overall diversity of test set.

Lines \texttt{9--13} describe the difficulty filter, where a fraction
of randomly selected ``easy'' phishing points, defined by highly confident
correct average scores predicted by three baseline models (Linear,
FFN, and GTE), are discarded. We use
$\delta = 0.15$, and $k = 10\%$.

Lines \text{16--25} describe the rate adjustment procedure, where for
each of five base rates, the number of phishing points ($n_p$)
and instances ($n_i$) are determined, and subsequently for
each base rate $n_i$ samples of size $n_p$ are sampled
without replacement from the reduced phishing test set.

\begin{algorithm}
\caption{Train-test split}\label{alg:test-set}
\begin{algorithmic}[1]
\State \textbf{Input:} $\mathcal{D}$ (clean dataset), $p$ (test set fraction), $\tau$ (similarity threshold)
\State \textbf{Output:} $\mathcal{D}_{train}$ (train set), $\mathcal{D}_{test}$ (test set)
\State
\State \textit{// temporal partition}
\State $\mathcal{D} \gets \texttt{SortByTime}(\mathcal{D})$
\State $\mathcal{D}_{train} \gets \{\mathcal{D}_{1:k} \mid k = \lfloor (1-p)* |\mathcal{D}|\rfloor \} $
\State $ \mathcal{D'}_{test} \gets \mathcal{D} \setminus  \mathcal{D}_{train} $
\State
\State \textit{// minimize leakage}
\State lsh $\gets$ \texttt{ConstructLSH}($\mathcal{D}_{train}$)
\State $\mathcal{D}_{test} \gets \emptyset$
\For{$x \in \mathcal{D'}_{test}$}
    \State candidates $\gets$ lsh.\texttt{search}($x$)
     \For{$c \in \text{candidates}$}
         \If{$\texttt{CosineDistance}(x, c) \geq \tau$}
            \State{$\mathcal{D}_{test} \gets \mathcal{D}_{test} \cup c$}
         \EndIf
     \EndFor
\EndFor
\State
\State \Return $\mathcal{D}_{train}, \mathcal{D}_{test}$
\end{algorithmic}
\end{algorithm}

\begin{algorithm}[h]
\caption{Benchmark dataset creation}\label{alg:benchmark-set}
\begin{algorithmic}[1]
\State \textbf{Input:} $\mathcal{D}_{test}$ (test dataset), $\mathcal{D}_{train}$ (train dataset) 
\State \textbf{Output:} {\bf $\mathcal{B}$} (benchmark datasets)
\State
\State \textit{// diversity enhancement}
\State $\mathcal{D}_{new} \gets \mathcal{D}_{test} \cup \mathcal{D}_{targetBenign} $
\State $\mathcal{D}_{new} \gets \mathcal{D}_{new} \setminus \mathcal{D}_{similarHTML} $
\State
\State \textit{// difficulty filter}
\State $ m \gets \text{fitModel}(\mathcal{D}_{train}) $
\State $ \text{score} \gets m.\text{predict}(\mathcal{D}_{new}) $
\State $ \mathcal{D}_{easy} \gets \{ x \mid \text{score}(x) > (1 - \delta) \wedge x \in P\} $ 
\State $ \mathcal{D}_{discard} \gets \text{sample}(k, \mathcal{D}_{easy}) $
\State $ \mathcal{D}_{new} \gets \mathcal{D}_{new} \setminus \mathcal{D}_{discard} $
\State
\State \textit{// base rate adjustment}
\State $\mathcal{D}_{benign} \gets \{ x \mid x \in \mathcal{D}_{test} \wedge x \in B \} $
\State $\mathcal{D}_{phish} \gets \{ x \mid x \in \mathcal{D}_{test} \wedge x \in P \} $ 
\State
\State $ \mathbf{\mathcal{B}} \gets \emptyset $
\For{$b \in \{0.05, 0.1, 0.5, 1, 5\}$} 
    \State  $n_p \gets \lfloor\frac{b \cdot |\mathcal{D}_{benign}|}{100 - b}\rfloor $ \Comment{num phish needed}
    \State $n_i \gets \texttt{GetInstances}(b)$ \Comment{num instances needed}
    \State $ \mathcal{D'}_{phish} \gets \texttt{RandomSample}(\mathcal{D}_{phish}, n_p, n_i, \texttt{replace=False}) $
    \State $ \mathbf{\mathcal{B}} \gets \mathbf{\mathcal{B}} \cup \{ \mathcal{D}_{benign}, \mathcal{D'}_{phish} \} $
\EndFor
\State
\State \Return $\mathcal{B}$
\end{algorithmic}
\end{algorithm}

\subsection{Dependence of Model Performance on Base Rate}
\label{sec:baseRateDependence}
  
The performance of a classifier degrades with decrease in the base
rate (fraction of the postive class) of a test data set.
Note that by model performance we mean precision (or variants) here.
While recall (true positive rate) and false positive rate (FPR) are
not affected by base rate (assuming a fixed, trained model), precision
is sensitive to it. In fact, one can write precision as a function of
recall, FPR, and base rate as:
$$ \text{Prec} = \frac{R \cdot r}{R \cdot r + \text{FPR} (1-r)} $$
 where $R$ is recall (or true positive rate), FPR is false positive rate and 
$r$ is base rate. It is easy to check that $ \frac{\partial \text{Prec}}{\partial r} $
is always positive, implying a monotonic, increasing relationship between precision and
base rate; so as base rate drops, precision tends to go down. Further, the above partial
derivative quantifies the sensitivity of precision to base rate.


\section{Baseline Experiments}
\label{sec:baseline-setup}

\subsection{Setup, Hyperparameters and Training Details}
The majority of experiments were performed on a single machine running
Ubuntu 20.04 with an Intel Xeon Gold 6230 CPU and 4x NVIDIA V100
GPUs, each with 32 GB VRAM.

\textbf{Linear model}
We implemented a linear support vector classifier (using LinearSVC in scikit-learn)
with squared hinge loss and L2 regularization.
Key hyperparameters included:
tolerance for convergence set to $10^{-3}$,
\texttt{dual=True} and \texttt{class\_weight='balanced'}.
The model used a combination of a few tens of hand-crafted features
and a large number of n-gram features.  The hand-crafted features
included URL-based, such as the overall length of the URL,
characteristics of the path, domain and subdomain structure, and
lengths of individual components; and those derived from the HTML DOM
structure, such as the number of nodes and edges in the DOM tree, to
capture the structural complexity of the web page.  The training
converged in 24 epochs.

\textbf{FFN model}
We used two types of n-gram features as input to
our feedforward neural network (FFN) model. The first type consists of
character-level 2-, 3-, and 4-grams extracted from the URL and
selected sections of the parsed HTML, including: visible text, HTML
tags, scripts, links, image sources, and embedded image data. The
second type of n-gram feature is derived from XPath sequences
extracted from the HTML DOM. Unlike the character-based n-grams, these
XPath n-grams are computed at the tag level. For example, the 2-gram
representation of the XPath sequence \texttt{<document>/\allowbreak<div>/\allowbreak<p>/\allowbreak<a>}
includes the segments: \texttt{<document>/\allowbreak<div>, <div>/\allowbreak<p>, and
  <p>/\allowbreak<a>}. Each character-level or XPath n-gram is treated as an
independent binary feature for input to the FFN.
The feature extraction process initially produced approximately 66.5
million binary features. We selected the 1.6 million most frequent
features from the training set for use in model training.
Training was conducted over 20 epochs. The entire
pipeline, including feature extraction and model training, completed
in approximately 45 hours.

\textbf{BERT-based model}
We fine-tuned \texttt{gte-large} pre-trained embedding model on our
training dataset using Hugging Face \texttt{trainer} with
\texttt{adamw} optimizer, a learning rate of $10^{-5}$,
momentum of 0.9 and a weight decay of $10^{-4}$. 
The model was fine-tuned for 2 epochs, taking approximately about 27 hours. 

\textbf{LLM-based prediction}
We used Google's \texttt{gemini-2.5-pro} model in batch mode to
classify webpages as phishing or benign based on the input URL and
HTML content. A system prompt was crafted to instruct the model to
assess social engineering indicators and output a JSON containing two
fields: a phishing flag (boolean) and a phishing risk score
(1--10). The HTML input is preprocessed
to remove non-essential elements and truncated to stay within token
limits. Each request includes the URL and simplified HTML in the
prompt, and the model returns structured output used for phishing
detection analysis. 


\textit{HTML simplification: }
To help shorten the input text, we remove unnecessary HTML elements such as metadata and styling.
Following guidance from prior work in phishing detection using LLMs \cite{koide2024chatphishdetector}, the simplification process consists of:
(1) Removing comments, \texttt{<style>}, and \texttt{<script>} tags; 
(2) Preserving content tags: \texttt{p}, \texttt{a}, \texttt{img}, \texttt{h1--h6}, \texttt{ul}, \texttt{ol}, \texttt{li};
(3) Filtering out tags with no text content;
(4) Normalizing links and images by stripping \texttt{http(s)://} and \texttt{www.};
(5) If token limit is exceeded despite above steps, HTML elements are trimmed from the center until token limit is met.

\textit{System prompt:}
\begin{Verbatim}[breaklines=true,breakanywhere=true,fontsize=\small,frame=single]
You are an expert in analyzing URL and multi-lingual HTML
to classify webpage as phishing or legitimate page.
Focus only on identifying if the page is phishing and
its corresponding risk score. Given the URL and HTML,
perform the following analysis for any Social Engineering
techniques often used in phishing attacks:

First perform thorough REASONING on the URL and HTML
looking for suspicious elements or inconsistencies,
domain issues, detect content red flags, evaluate
security indicators and inconsistencies, and then
classify with a confidence score.

1.  **PHISHING:** Whether the webpage is a phishing page
    or a legitimate page. (True if phishing, False if
    legitimate)
2.  **SCORE:** Indicates phishing risk on a scale of
    0.0 to 1.0 (inclusive), where 0.0 is the least likely
    to have phishing content and 1.0 is the most likely.
3.  **REASON:** Briefly (one sentence) explain the
    reasoning behind your determination.

Provide the extracted information in JSON format with the
following keys:
  - phishing: boolean (whether the site is a phishing site
    or a legitimate site)
  - score: float (indicates phishing risk on a scale of 0.0
    to 1.0)
  - reason: str (one sentence reason)
\end{Verbatim}

We note that Gemini refused to score 52 samples (46 benign and 6 phishing)
due to containing prohibited contents. These instances were exceluded from
the analysis when computing the performance of this model.

\subsection{Benchmark evaluation results}
We provide several visualizations to show the performance of the
four models on the full (24\%) benchmark dataset. Similar plots
are also shown for the test set in \ref{sec:test-results}

\begin{figure*}[t]
  \centering
  \begin{adjustbox}{max height=0.47\textheight, center}
  \begin{minipage}{\textwidth}
    \centering
    \begin{subfigure}{\textwidth}
      \centering
      \begin{minipage}[t]{0.48\textwidth}
        \centering
        \includegraphics[width=\linewidth]{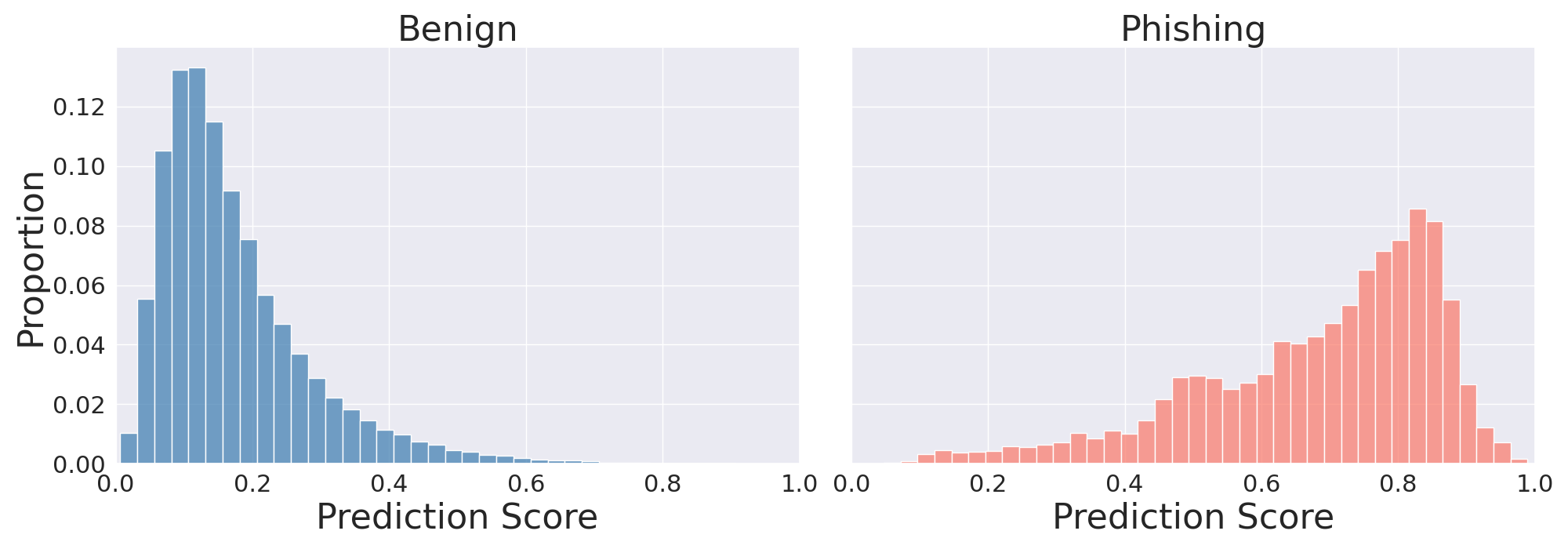}
      \end{minipage}%
      \hfill
      \begin{minipage}[t]{0.48\textwidth}
        \centering
        \includegraphics[width=\linewidth]{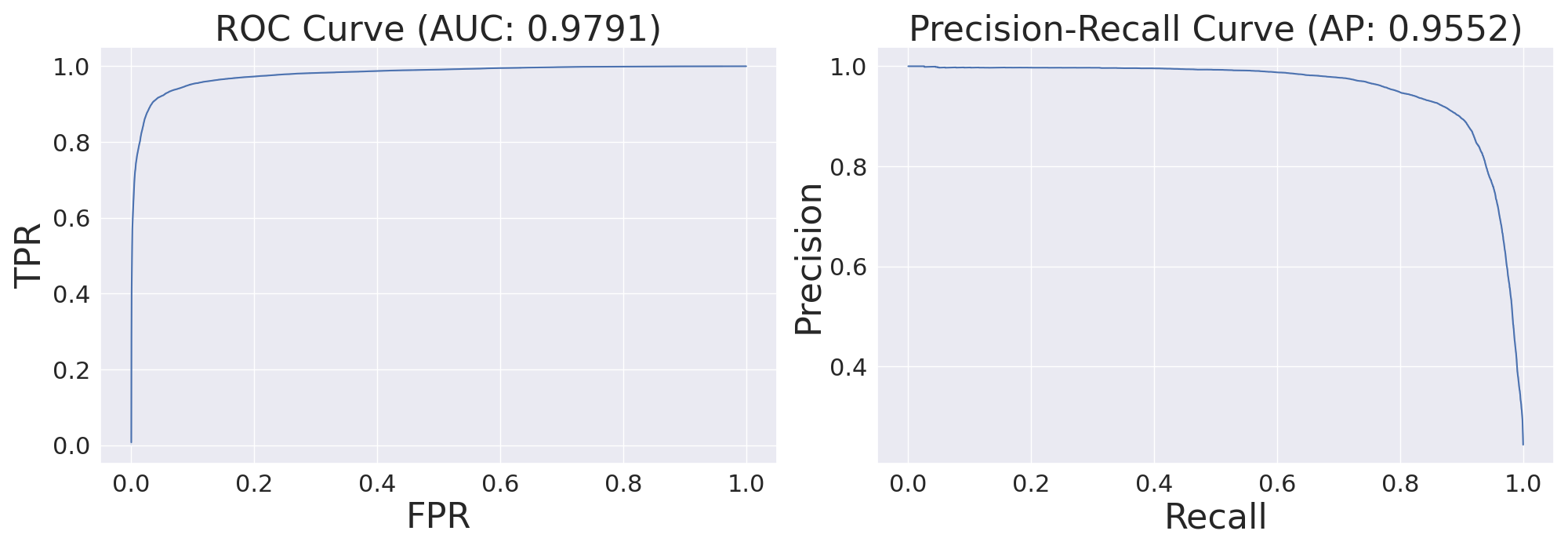}
      \end{minipage}
      \caption{Linear}
      \label{fig:bench-linear}
    \end{subfigure}
    \vspace{4pt}
    
    \begin{subfigure}{\textwidth}
      \centering
      \begin{minipage}[t]{0.48\textwidth}
        \centering
        \includegraphics[width=\linewidth]{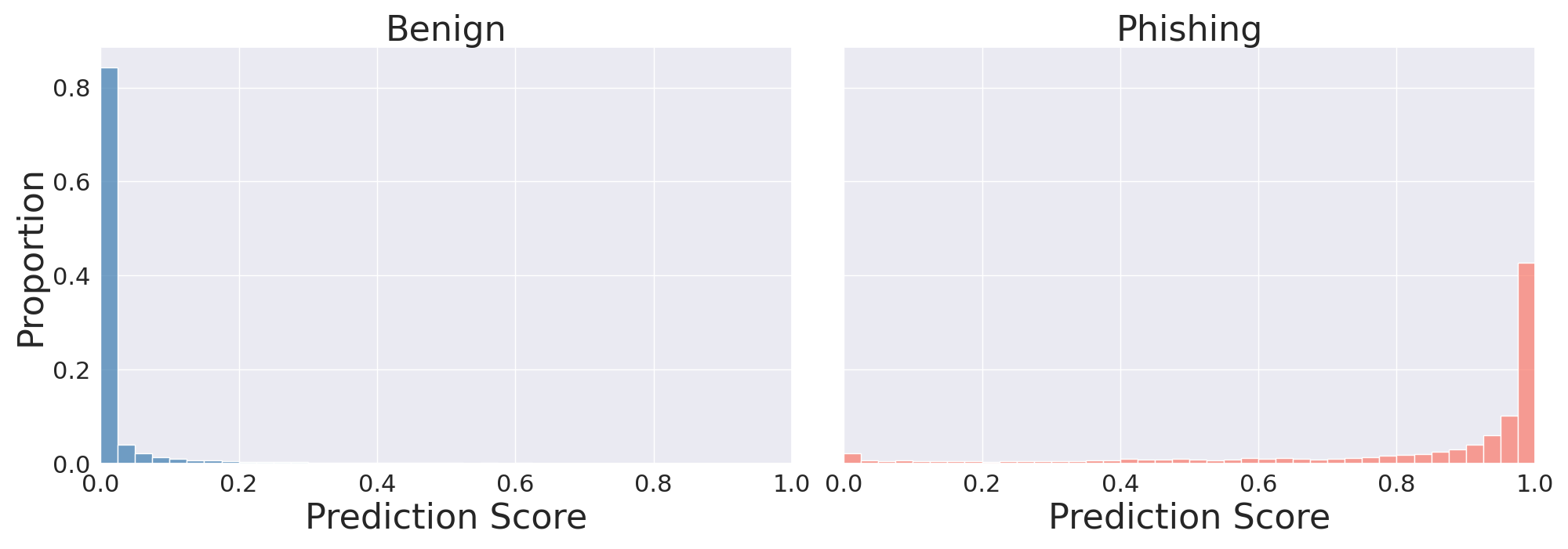}
      \end{minipage}%
      \hfill
      \begin{minipage}[t]{0.48\textwidth}
        \centering
        \includegraphics[width=\linewidth]{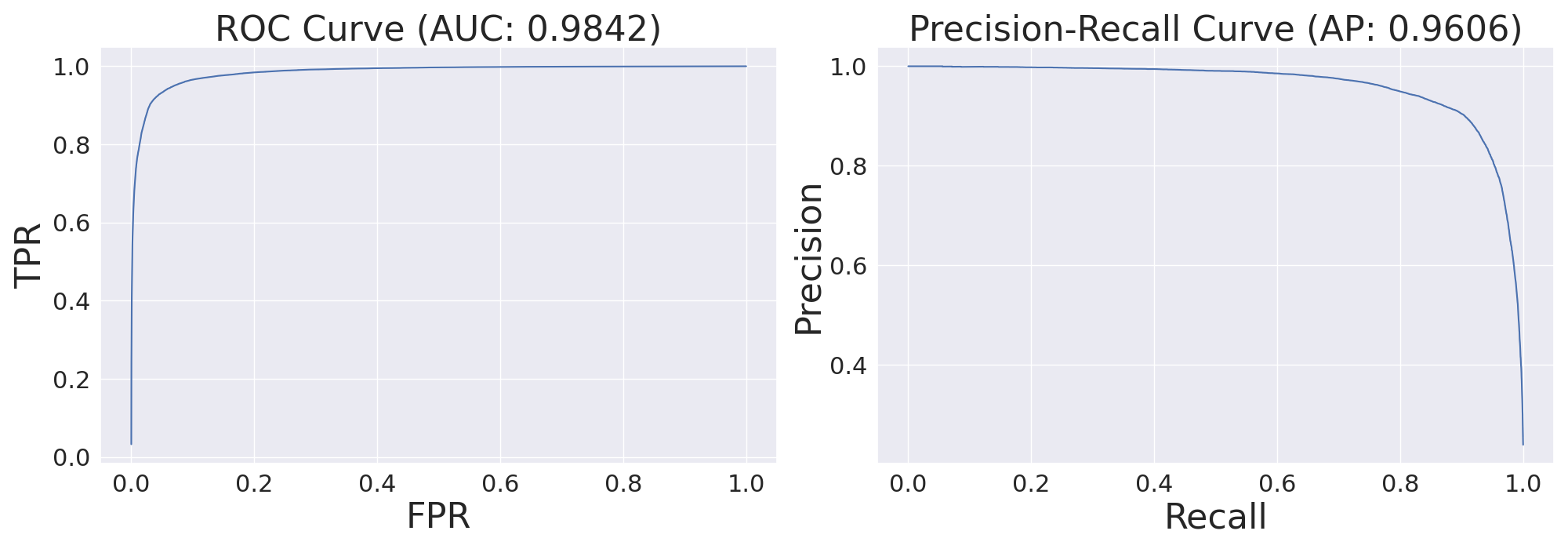}
      \end{minipage}
      \caption{FFN}
      \label{fig:bench-ffn}
    \end{subfigure}
    \vspace{4pt}
    
    \begin{subfigure}{\textwidth}
      \centering
      \begin{minipage}[t]{0.48\textwidth}
        \centering
        \includegraphics[width=\linewidth]{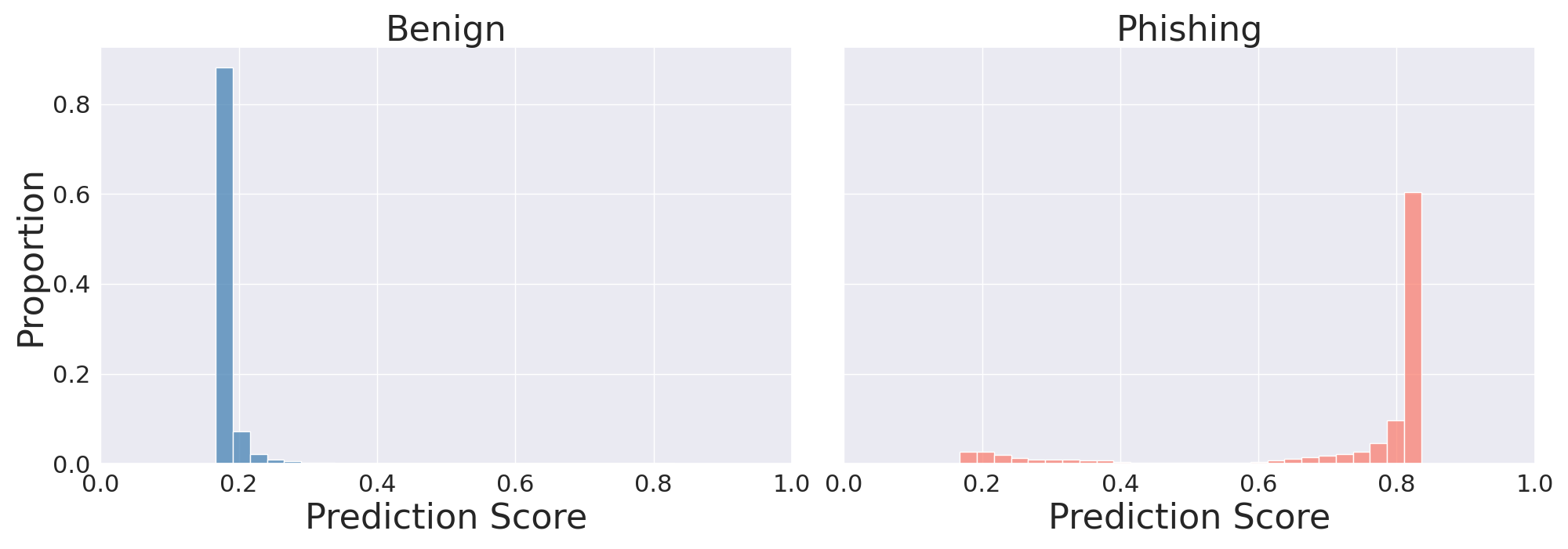}
      \end{minipage}%
      \hfill
      \begin{minipage}[t]{0.48\textwidth}
        \centering
        \includegraphics[width=\linewidth]{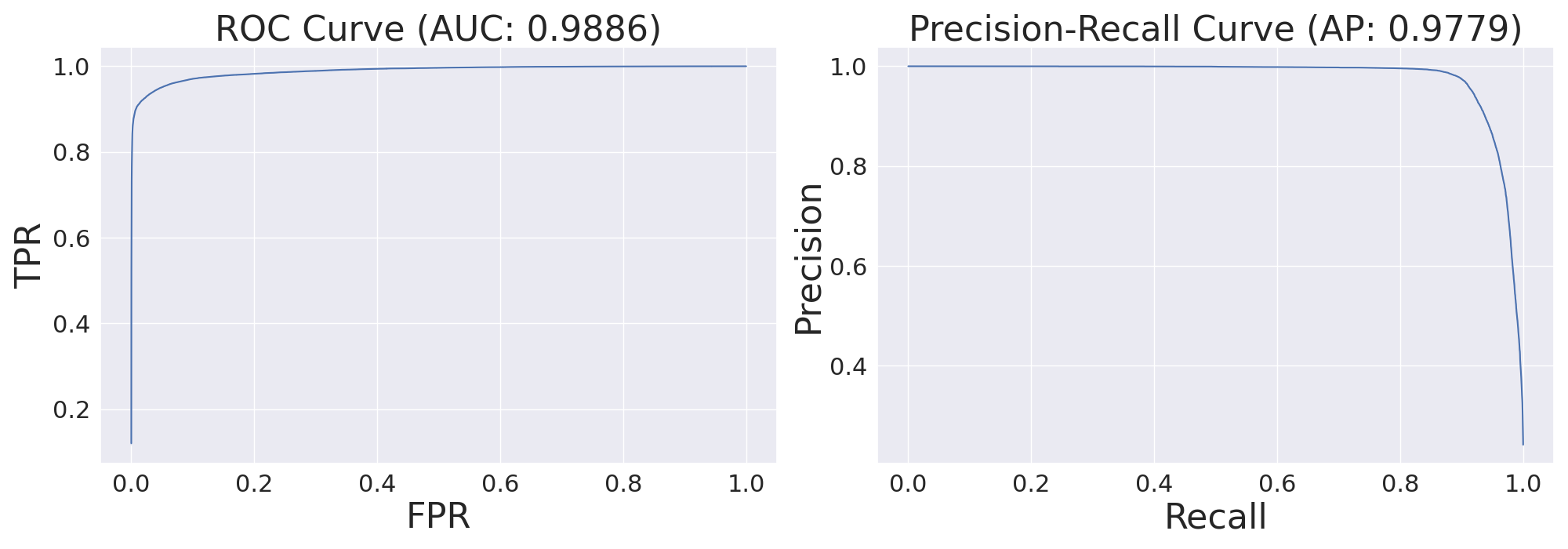}
      \end{minipage}
      \caption{GTE}
      \label{fig:bench-gte}
    \end{subfigure}
    \vspace{4pt}
    
    \begin{subfigure}{\textwidth}
      \centering
      \begin{minipage}[t]{0.48\textwidth}
        \centering
        \includegraphics[width=\linewidth]{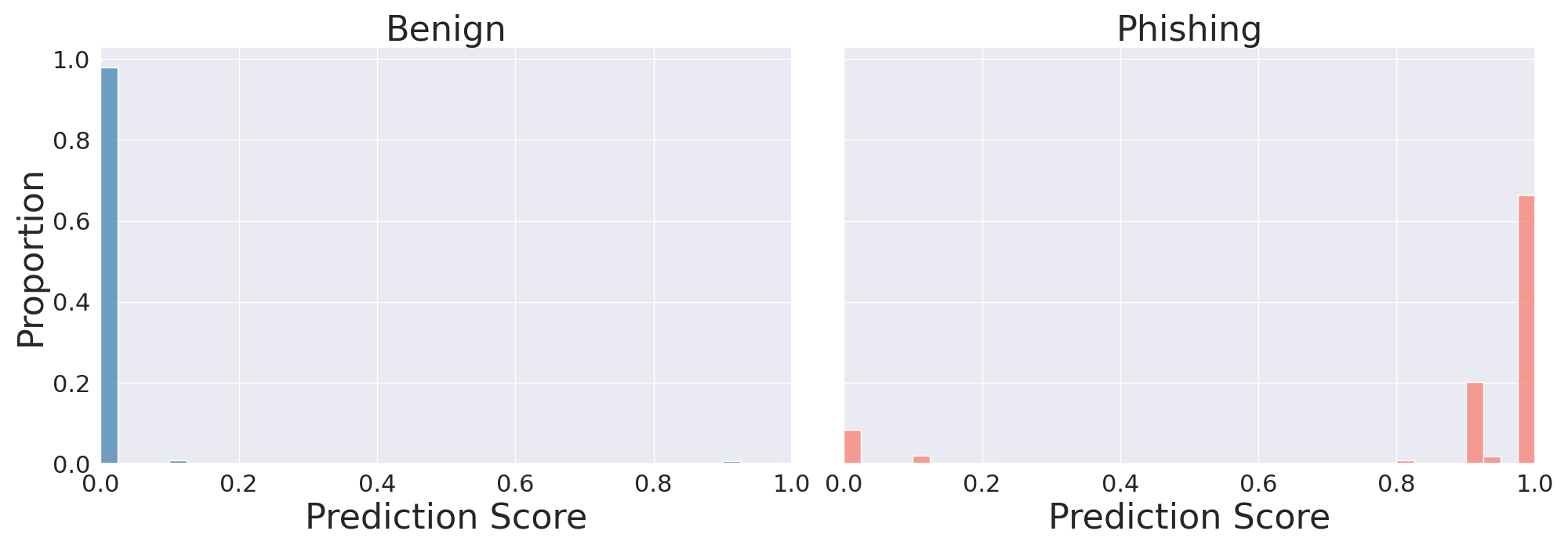}
      \end{minipage}%
      \hfill
      \begin{minipage}[t]{0.48\textwidth}
        \centering
        \includegraphics[width=\linewidth]{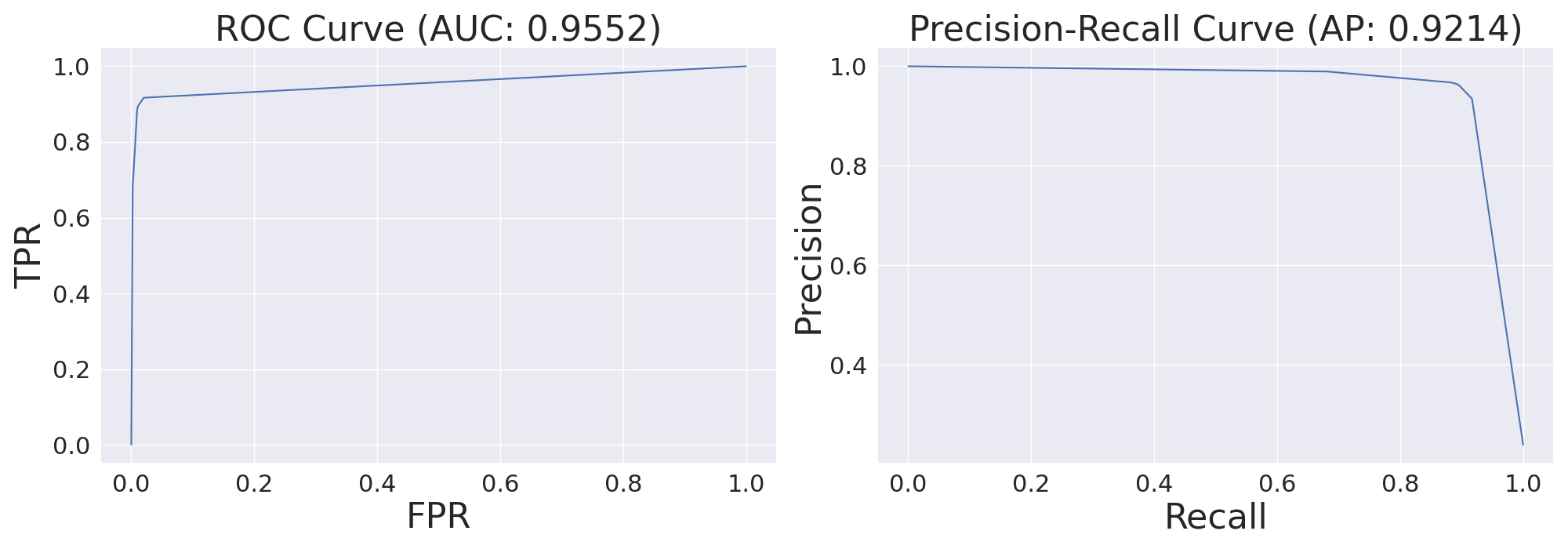}
      \end{minipage}
      \caption{LLM}
      \label{fig:bench-llm}
    \end{subfigure}
  \end{minipage}
  \end{adjustbox}
  \caption{Model performance on the full (24\%) benchmark dataset: (left) score distributions, (right) ROC and PR curves.}
  \label{fig:bench-score-curves}
\end{figure*}

\textbf{Prediction score histograms}
Figure~\ref{fig:bench-score-curves} (left) shows the score histogram on the benchmark set 
for each of the models, with separate plots for benign and phishing points. 
The GTE model which has the best performance based on average
precision also shows good class separation, with minimal overlap in
the central score region. Further, the sharp peaks near zero and one,
for the benign and phishing points, respectively, shows high
confidence in correctly classified instances.  Other models show
multimodal distributions for the two classes, with lesser separation
and greater overlap in the mid-score region.

\textbf{ROC and PR curves}
Figure~\ref{fig:bench-score-curves} (right) presents the ROC and precision-recall
(PR) curves for the benchmark dataset.  Most models show strong
performance, with the exception of the LLM-based model.  The ROC-AUC
values are quite close to 1.0, since the FPR values are numerically
low, even at 24\% base rate.  As the base rate decreases, the FPR for
a given level of precision tends to become even smaller.
Consequently, we omit ROC curves for lower base rates.

\begin{figure*}[t]
  \centering

  \begin{subfigure}[b]{\textwidth}
    \centering
    \includegraphics[width=\linewidth]{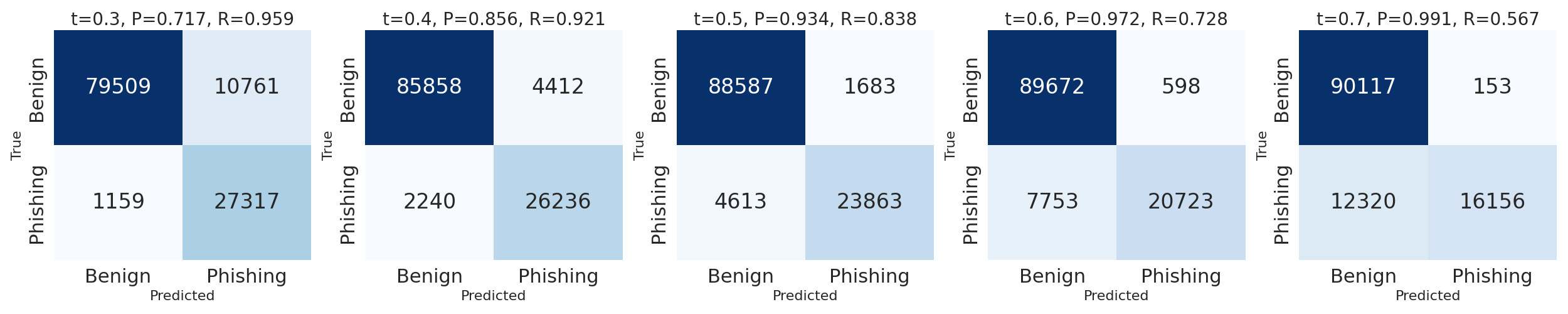}
    \caption{Linear}
    \label{fig:bench-linear-confusion}
  \end{subfigure}

  \vfill

  \begin{subfigure}[b]{\textwidth}
    \centering
    \includegraphics[width=\linewidth]{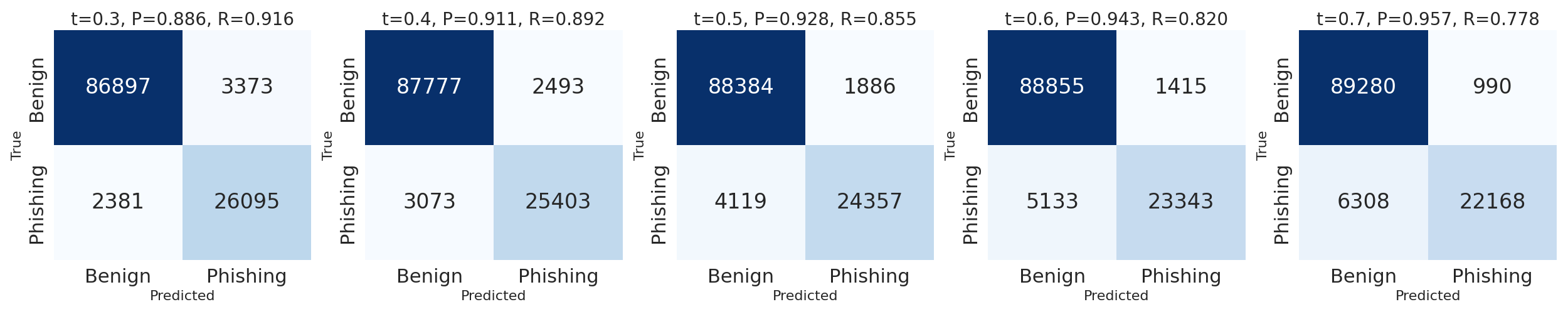}
    \caption{FFN}
    \label{fig:bench-ffn-confusion}
  \end{subfigure}

  \vfill

  \begin{subfigure}[b]{\textwidth}
    \centering
    \includegraphics[width=\linewidth]{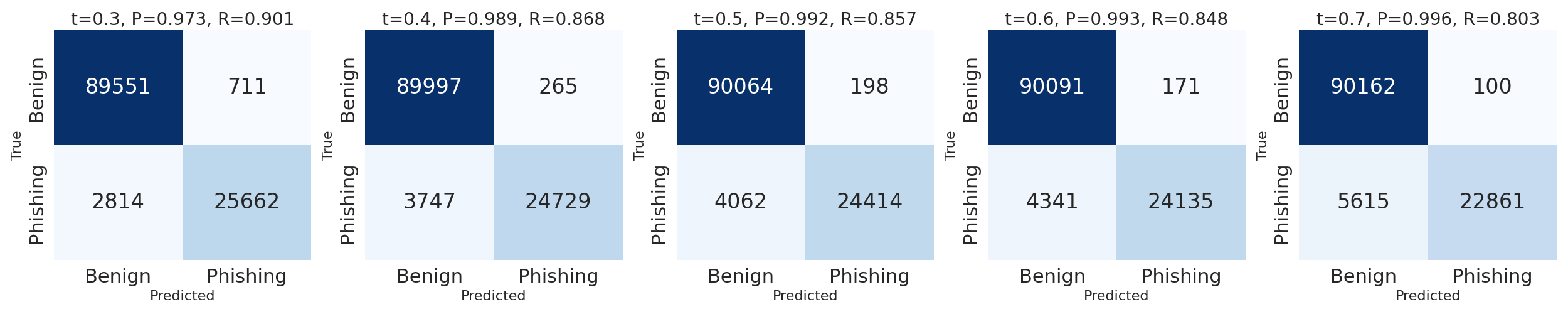}
    \caption{GTE}
    \label{fig:bench-gte-confusion}
  \end{subfigure}

  \vfill

  \begin{subfigure}[b]{\textwidth}
    \centering
    \includegraphics[width=\linewidth]{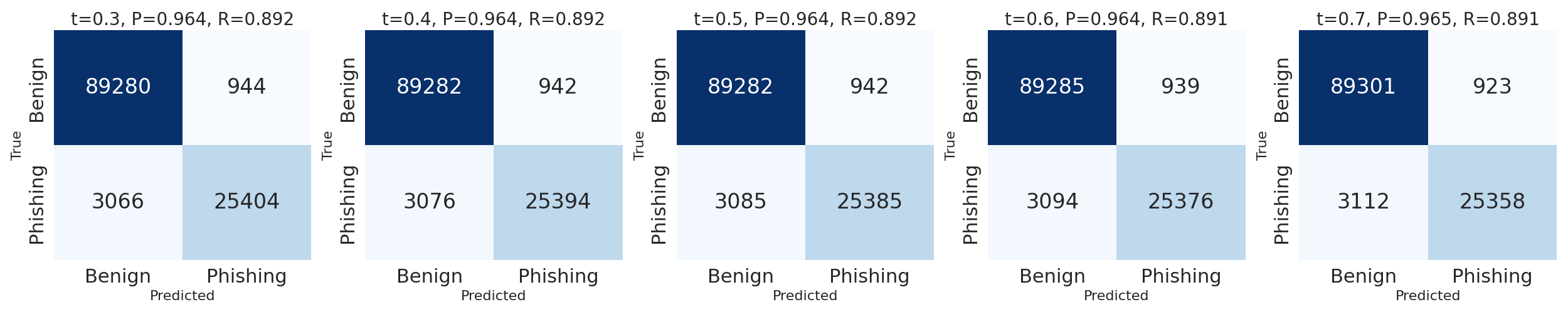}
    \caption{LLM}
    \label{fig:bench-llm-confusion}
  \end{subfigure}

  \caption{Confusion matrices at different thresholds (t) for the full (24\%) benchmark dataset. In addition, the precision (P) and recall (R) at that threshold is also reported.}
  \label{fig:bench-confusion}
\end{figure*}

\textbf{Confusion matrices}
Figure~\ref{fig:bench-confusion} shows the trade off between precision
and recall as the model threshold is changed.
As the threshold increases the recall decreases while the precision
increases. For example, at threhold of 0.7, the GTE model achieves a
precision of 99.6\% at a recall of 80.3\%.

\subsection{Test evaluation results}
\label{sec:test-results}

\begin{figure*}[t]
  \centering
  \begin{adjustbox}{max height=0.47\textheight, center}
  \begin{minipage}{\textwidth}
    \centering
    \begin{subfigure}{\textwidth}
      \centering
      \begin{minipage}[t]{0.48\textwidth}
        \centering
        \includegraphics[width=\linewidth]{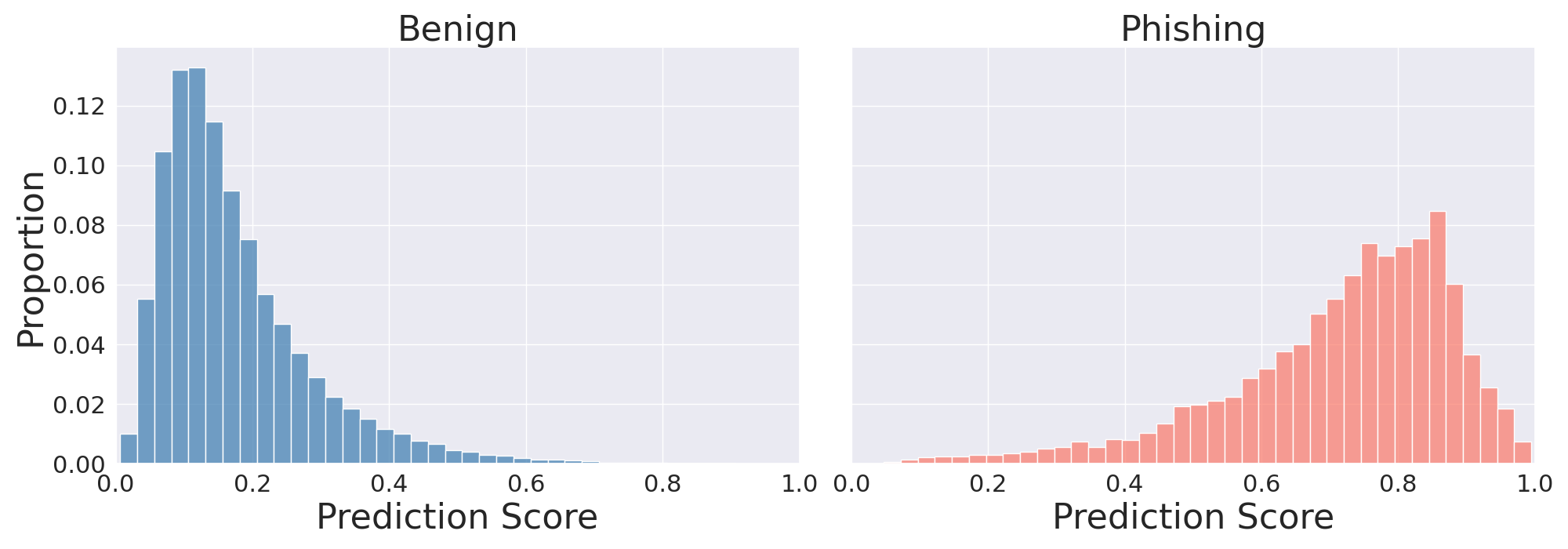}
      \end{minipage}%
      \hfill
      \begin{minipage}[t]{0.48\textwidth}
        \centering
        \includegraphics[width=\linewidth]{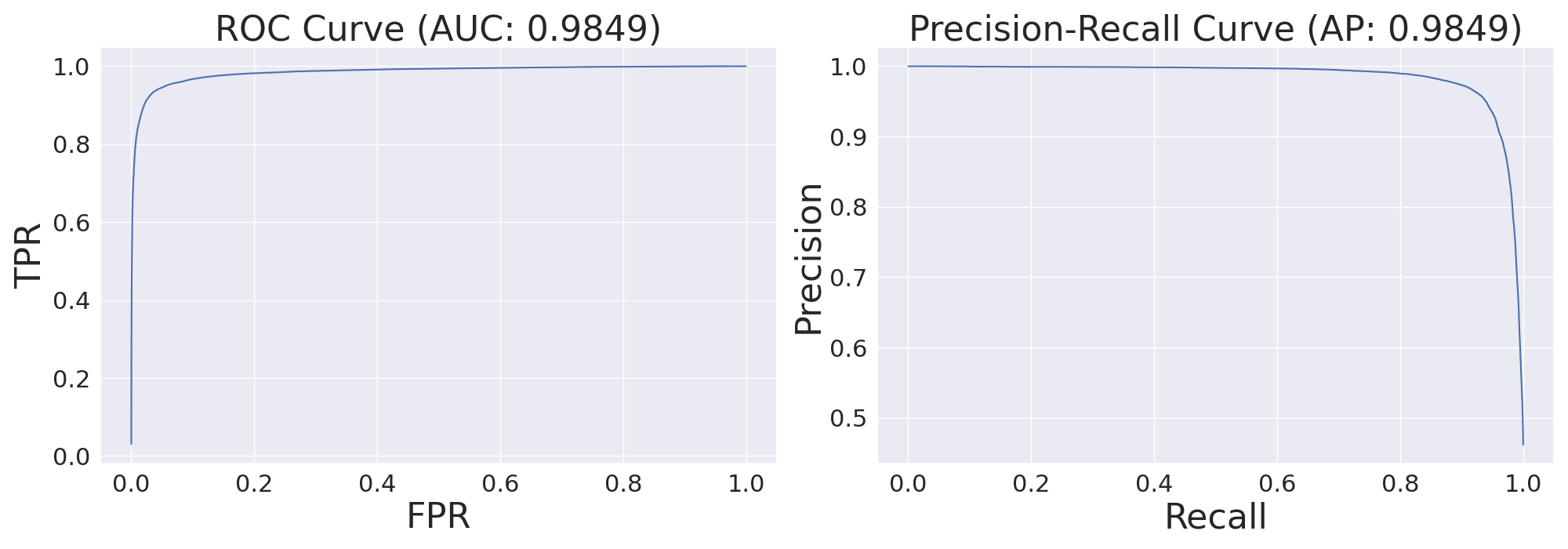}
      \end{minipage}
      \caption{Linear}
      \label{fig:test-linear}
    \end{subfigure}
    \vspace{4pt}
    
    \begin{subfigure}{\textwidth}
      \centering
      \begin{minipage}[t]{0.48\textwidth}
        \centering
        \includegraphics[width=\linewidth]{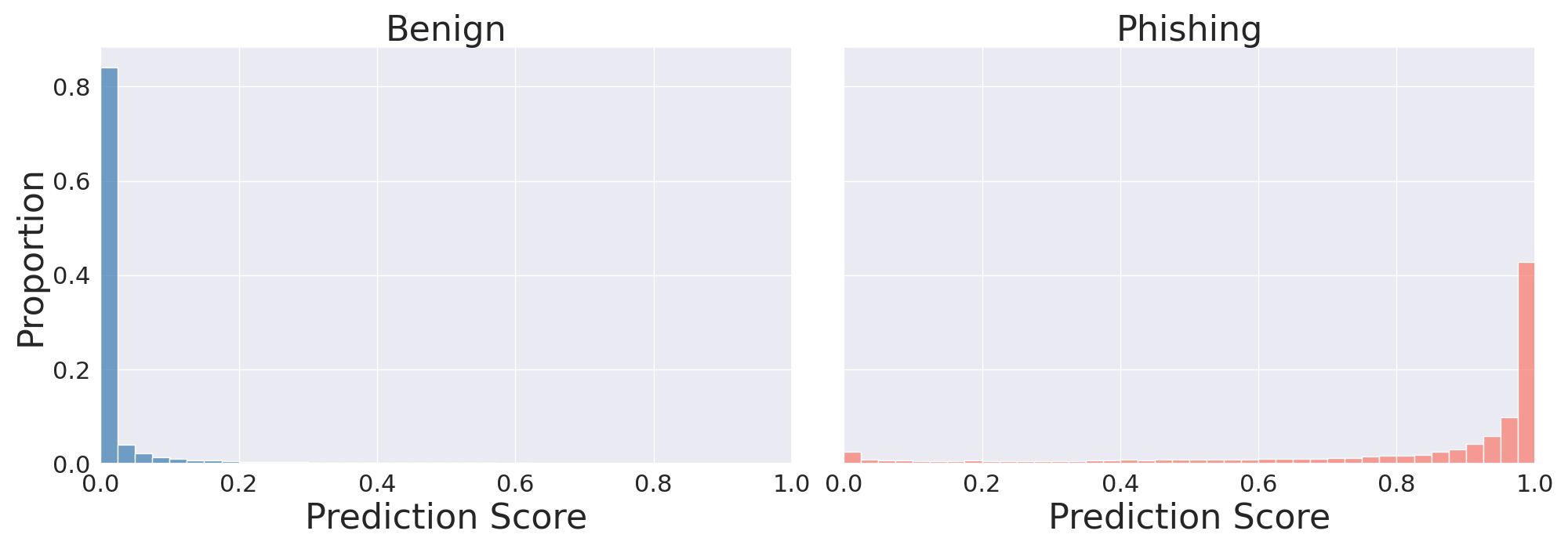}
      \end{minipage}%
      \hfill
      \begin{minipage}[t]{0.48\textwidth}
        \centering
        \includegraphics[width=\linewidth]{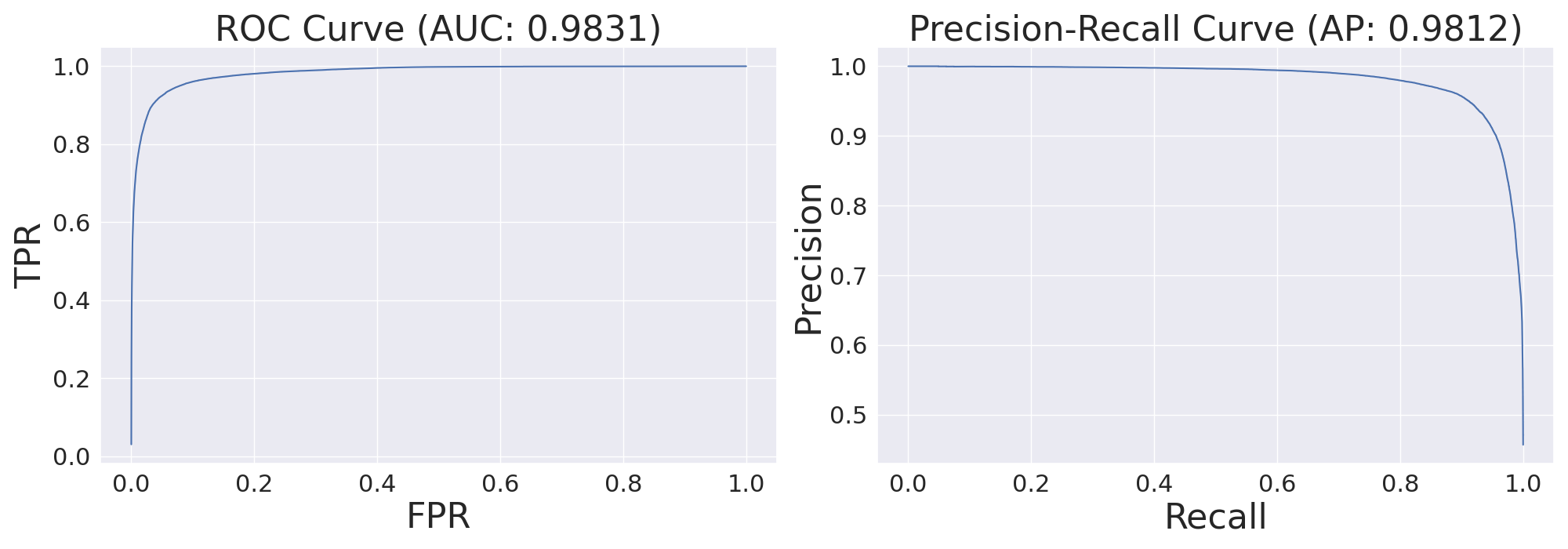}
      \end{minipage}
      \caption{FFN}
      \label{fig:test-ffn}
    \end{subfigure}
    \vspace{4pt}
    
    \begin{subfigure}{\textwidth}
      \centering
      \begin{minipage}[t]{0.48\textwidth}
        \centering
        \includegraphics[width=\linewidth]{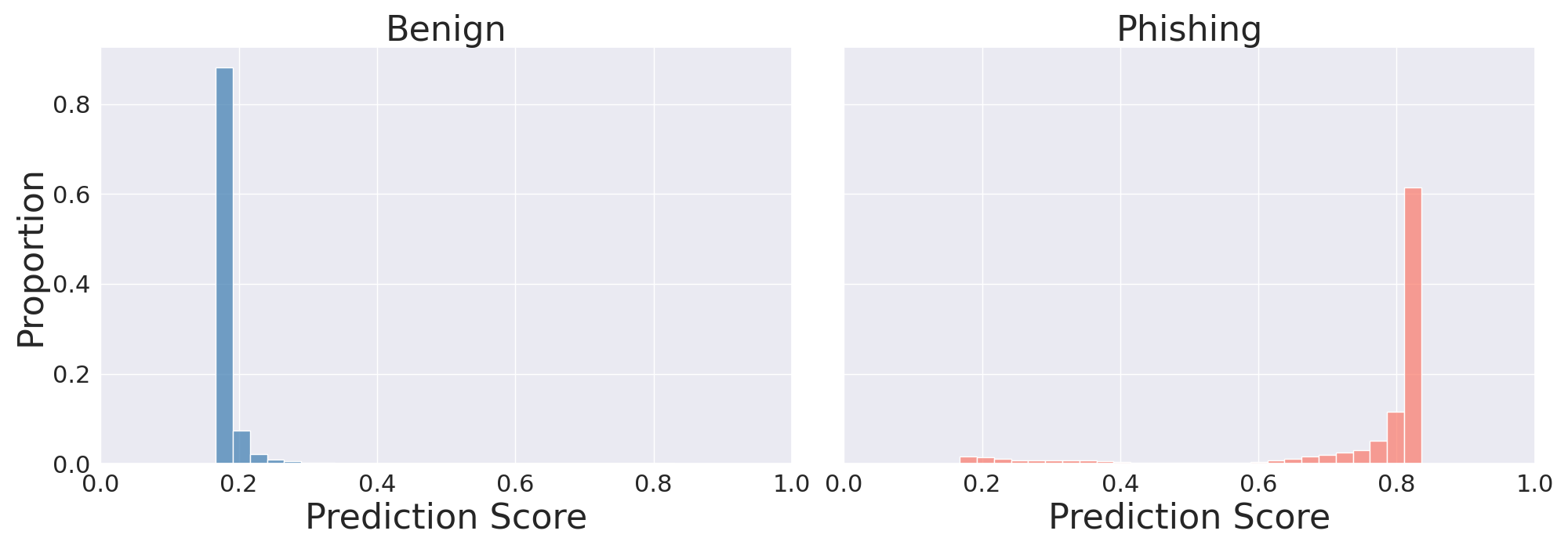}
      \end{minipage}%
      \hfill
      \begin{minipage}[t]{0.48\textwidth}
        \centering
        \includegraphics[width=\linewidth]{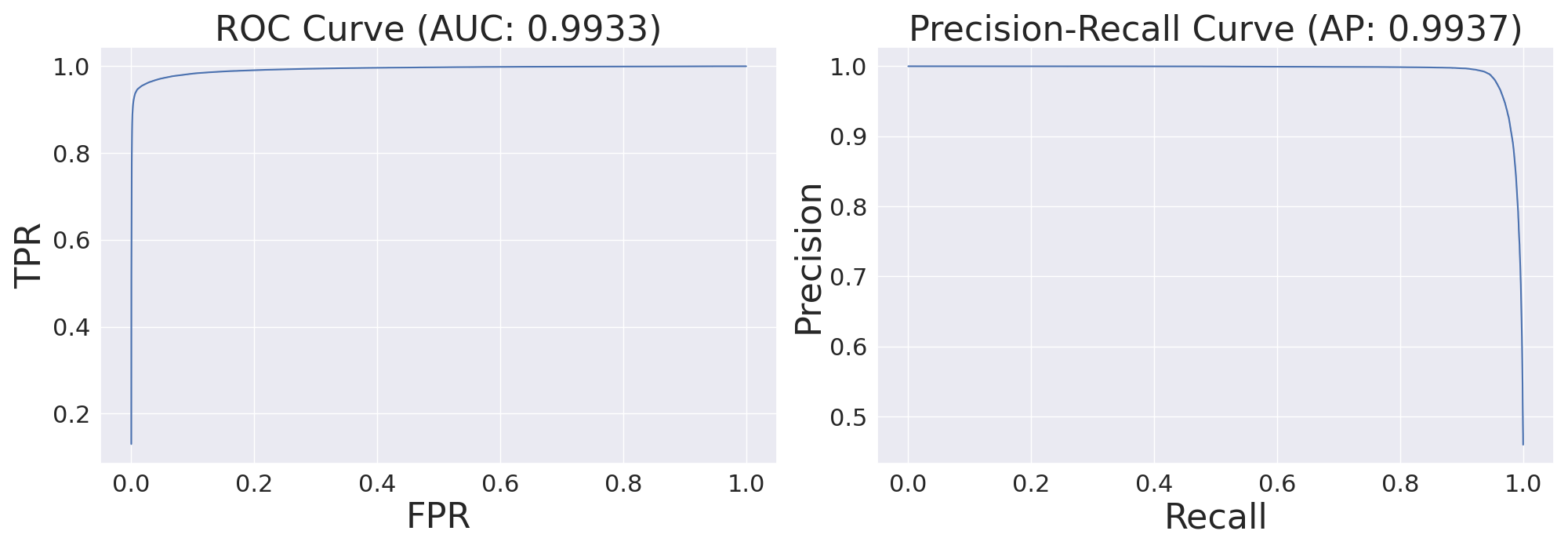}
      \end{minipage}
      \caption{GTE}
      \label{fig:test-gte}
    \end{subfigure}
    \vspace{4pt}
    
    \begin{subfigure}{\textwidth}
      \centering
      \begin{minipage}[t]{0.48\textwidth}
        \centering
        \includegraphics[width=\linewidth]{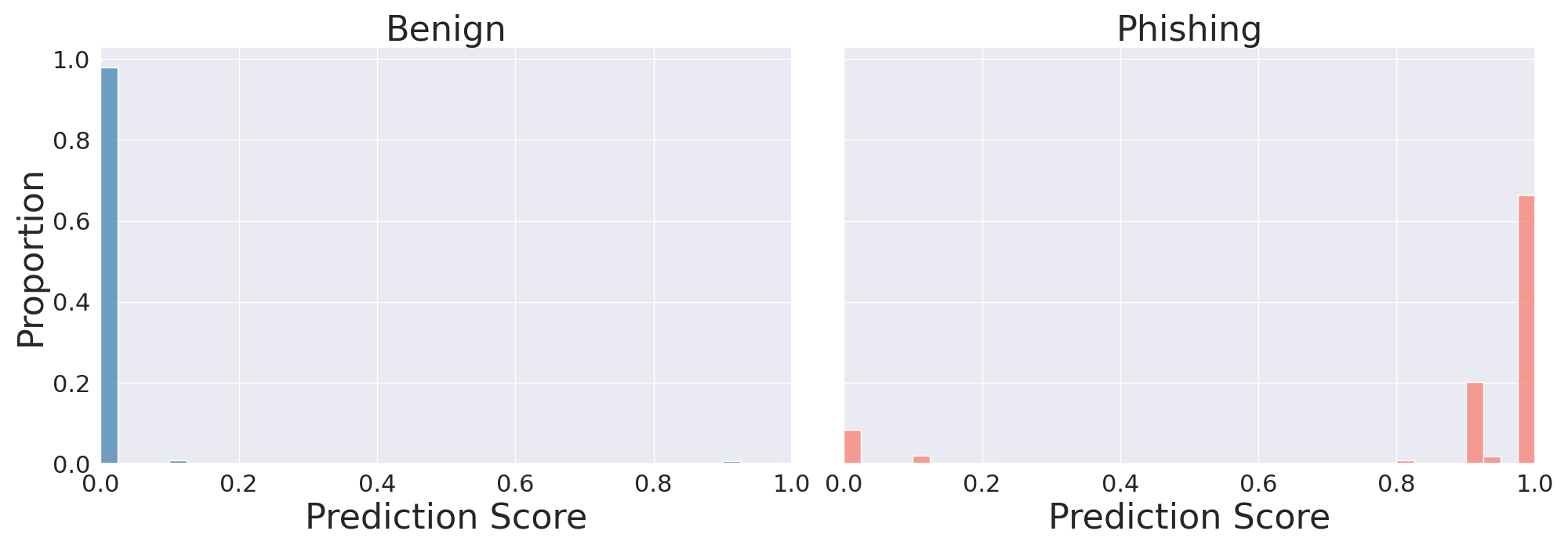}
      \end{minipage}%
      \hfill
      \begin{minipage}[t]{0.48\textwidth}
        \centering
        \includegraphics[width=\linewidth]{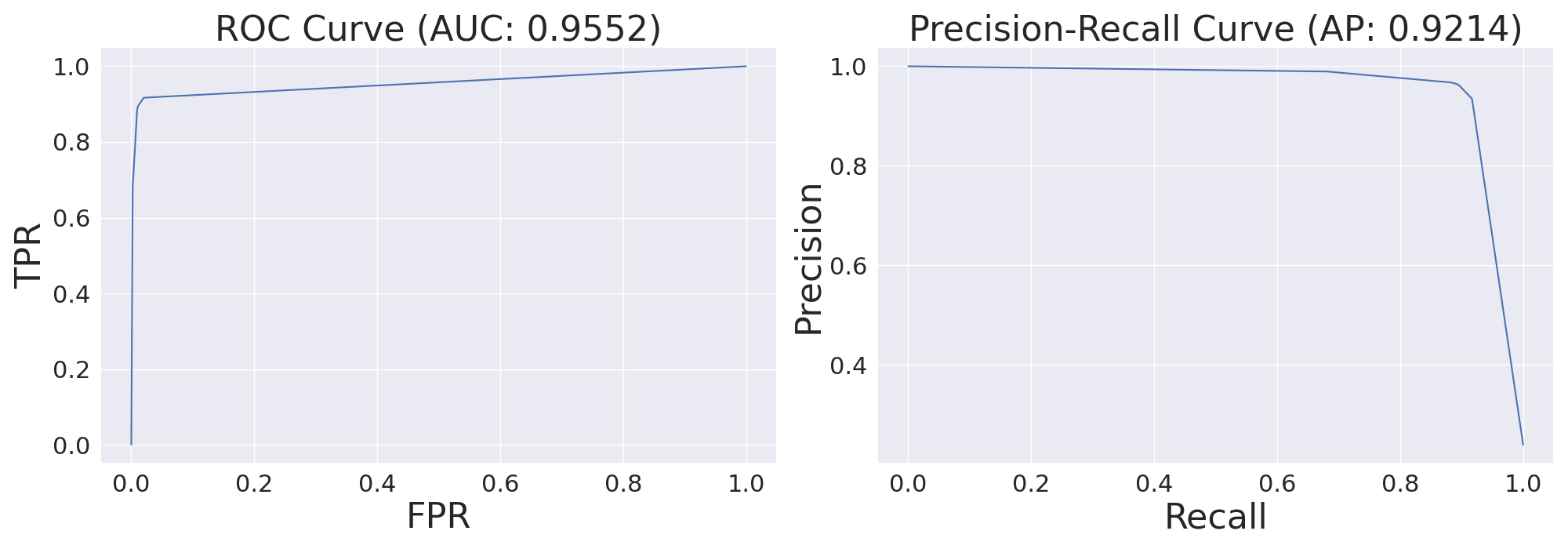}
      \end{minipage}
      \caption{LLM}
      \label{fig:test-llm}
    \end{subfigure}
  \end{minipage}
  \end{adjustbox}
  \caption{Model performance on the test dataset: (left) score distributions, (right) ROC and PR curves.}
  \label{fig:test-score-curves}
\end{figure*}

\begin{figure*}[t]
  \centering

  \begin{subfigure}[b]{\textwidth}
    \centering
    \includegraphics[width=\linewidth]{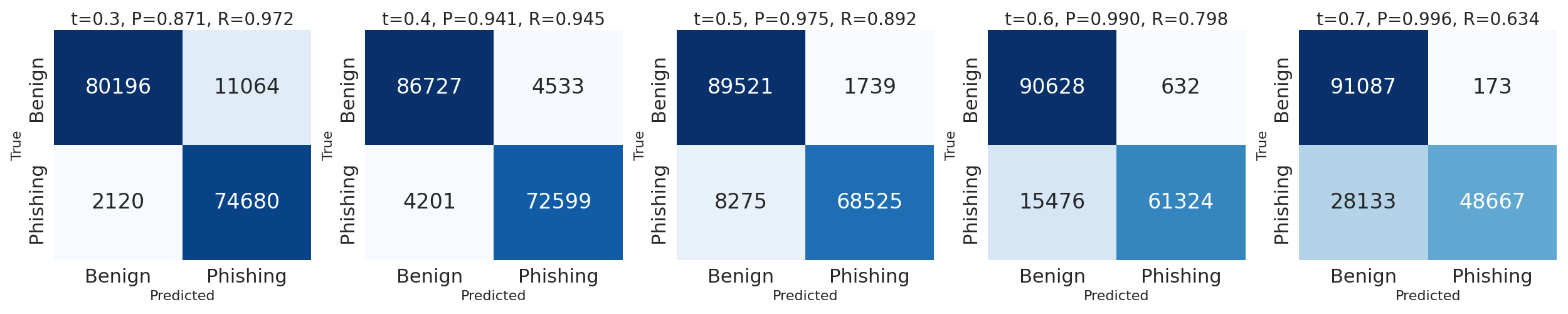}
    \caption{Linear}
    \label{fig:test-linear-confusion}
  \end{subfigure}

  \vfill

  \begin{subfigure}[b]{\textwidth}
    \centering
    \includegraphics[width=\linewidth]{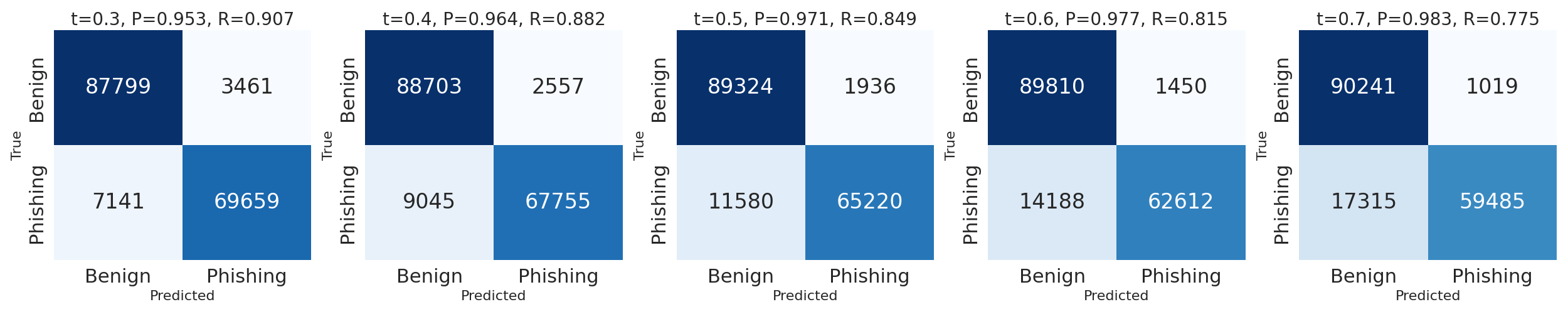}
    \caption{FFN}
    \label{fig:test-ffn-confusion}
  \end{subfigure}

  \vfill

  \begin{subfigure}[b]{\textwidth}
    \centering
    \includegraphics[width=\linewidth]{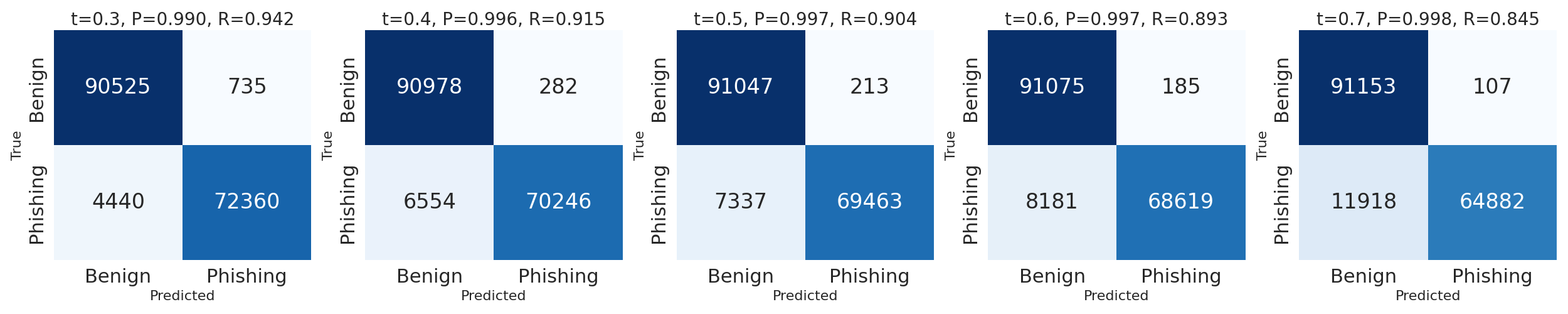}
    \caption{GTE}
    \label{fig:test-gte-confusion}
  \end{subfigure}

  \vfill

  \begin{subfigure}[b]{\textwidth}
    \centering
    \includegraphics[width=\linewidth]{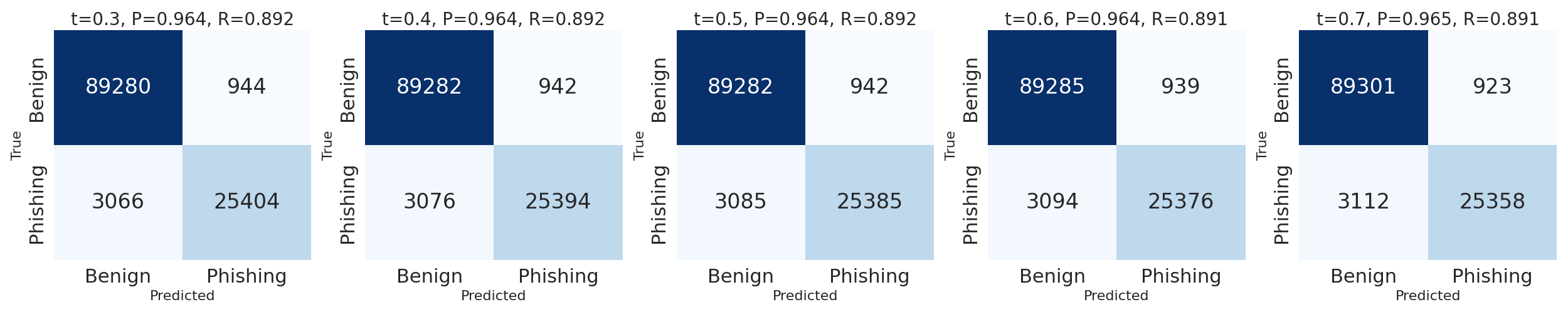}
    \caption{LLM}
    \label{fig:test-llm-confusion}
  \end{subfigure}

  \caption{Confusion matrices at different thresholds (t) for the test dataset. In addition, the precision (P) and recall (R) at that threshold is also reported.}
  \label{fig:test-confusion}
\end{figure*}

\end{document}